\PassOptionsToPackage{unicode}{hyperref}
\PassOptionsToPackage{hyphens}{url}
\PassOptionsToPackage{dvipsnames,svgnames,x11names}{xcolor}
\documentclass[
  12pt,
  a4paper,
]{article}
\usepackage{xcolor}
\usepackage[margin=1in]{geometry}
\usepackage{amsmath,amssymb}
\usepackage{iftex}
\ifPDFTeX
  \usepackage[T1]{fontenc}
  \usepackage[utf8]{inputenc}
  \usepackage{textcomp} % provide euro and other symbols
\else % if luatex or xetex
  \usepackage{unicode-math} % this also loads fontspec
  \defaultfontfeatures{Scale=MatchLowercase}
  \defaultfontfeatures[\rmfamily]{Ligatures=TeX,Scale=1}
\fi
\usepackage{lmodern}
\ifPDFTeX\else
\fi
\IfFileExists{upquote.sty}{\usepackage{upquote}}{}
\IfFileExists{microtype.sty}{% use microtype if available
  \usepackage[]{microtype}
  \UseMicrotypeSet[protrusion]{basicmath} % disable protrusion for tt fonts
}{}
\usepackage{setspace}
\makeatletter
\@ifundefined{KOMAClassName}{% if non-KOMA class
  \IfFileExists{parskip.sty}{%
    \usepackage{parskip}
  }{% else
    \setlength{\parindent}{0pt}
    \setlength{\parskip}{6pt plus 2pt minus 1pt}}
}{% if KOMA class
  \KOMAoptions{parskip=half}}
\makeatother
\makeatletter
\ifx\paragraph\undefined\else
  \let\oldparagraph\paragraph
  \renewcommand{\paragraph}{
    \@ifstar
      \xxxParagraphStar
      \xxxParagraphNoStar
  }
  \newcommand{\xxxParagraphStar}[1]{\oldparagraph*{#1}\mbox{}}
  \newcommand{\xxxParagraphNoStar}[1]{\oldparagraph{#1}\mbox{}}
\fi
\ifx\subparagraph\undefined\else
  \let\oldsubparagraph\subparagraph
  \renewcommand{\subparagraph}{
    \@ifstar
      \xxxSubParagraphStar
      \xxxSubParagraphNoStar
  }
  \newcommand{\xxxSubParagraphStar}[1]{\oldsubparagraph*{#1}\mbox{}}
  \newcommand{\xxxSubParagraphNoStar}[1]{\oldsubparagraph{#1}\mbox{}}
\fi
\makeatother

\usepackage{longtable,booktabs,array}
\usepackage{calc} % for calculating minipage widths
\usepackage{etoolbox}
\makeatletter
\patchcmd\longtable{\par}{\if@noskipsec\mbox{}\fi\par}{}{}
\makeatother
\IfFileExists{footnotehyper.sty}{\usepackage{footnotehyper}}{\usepackage{footnote}}
\makesavenoteenv{longtable}
\usepackage{graphicx}
\makeatletter
\newsavebox\pandoc@box
\newcommand*\pandocbounded[1]{% scales image to fit in text height/width
  \sbox\pandoc@box{#1}%
  \Gscale@div\@tempa{\textheight}{\dimexpr\ht\pandoc@box+\dp\pandoc@box\relax}%
  \Gscale@div\@tempb{\linewidth}{\wd\pandoc@box}%
  \ifdim\@tempb\p@<\@tempa\p@\let\@tempa\@tempb\fi% select the smaller of both
  \ifdim\@tempa\p@<\p@\scalebox{\@tempa}{\usebox\pandoc@box}%
  \else\usebox{\pandoc@box}%
  \fi%
}
\def\fps@figure{htbp}
\makeatother

\providecommand{\tightlist}{%
  \setlength{\itemsep}{0pt}\setlength{\parskip}{0pt}}

\usepackage{algorithm}
\usepackage{algpseudocode}
\usepackage{float}
 
\usepackage[style=apa,backend=biber,sorting=nyt,doi=true,url=true,isbn=false,uniquename=false,uniquelist=false,natbib=true]{biblatex}
\usepackage{authblk}

\usepackage[tablesonly,nolists]{endfloat}
\usepackage{placeins}
\usepackage{afterpage}
\makeatletter
\@ifpackageloaded{caption}{}{\usepackage{caption}}
\AtBeginDocument{%
\ifdefined\contentsname
  \renewcommand*\contentsname{Table of contents}
\else
  \newcommand\contentsname{Table of contents}
\fi
\ifdefined\listfigurename
  \renewcommand*\listfigurename{List of Figures}
\else
  \newcommand\listfigurename{List of Figures}
\fi
\ifdefined\listtablename
  \renewcommand*\listtablename{List of Tables}
\else
  \newcommand\listtablename{List of Tables}
\fi
\ifdefined\figurename
  \renewcommand*\figurename{Figure}
\else
  \newcommand\figurename{Figure}
\fi
\ifdefined\tablename
  \renewcommand*\tablename{Table}
\else
  \newcommand\tablename{Table}
\fi
}
\@ifpackageloaded{float}{}{\usepackage{float}}
\floatstyle{ruled}
\@ifundefined{c@chapter}{\newfloat{codelisting}{h}{lop}}{\newfloat{codelisting}{h}{lop}[chapter]}
\floatname{codelisting}{Listing}

\usepackage{amsthm}
\theoremstyle{definition}
\newtheorem{example}{Example}[section]
\theoremstyle{plain}
\newtheorem{lemma}{Lemma}[section]
\theoremstyle{plain}
\newtheorem{proposition}{Proposition}[section]
\theoremstyle{definition}
\newtheorem{definition}{Definition}[section]
\theoremstyle{remark}
\AtBeginDocument{}
\newtheorem*{remark}{Remark}

\makeatother
\makeatletter
\@ifpackageloaded{caption}{}{\usepackage{caption}}
\@ifpackageloaded{subcaption}{}{\usepackage{subcaption}}
\makeatother
\usepackage{bookmark}
\IfFileExists{xurl.sty}{\usepackage{xurl}}{} % add URL line breaks if available
\hypersetup{
  pdftitle={Informative Distance-Based Priors for Correlation Matrices Centred on a Target Reference},
  pdfkeywords={Correlation matrix, Penalised Complexity prior, LKJ
prior, Cholesky factorisation, Gaussian Markov random field},
  colorlinks=true,
  linkcolor={blue},
  filecolor={Maroon},
  citecolor={Blue},
  urlcolor={Blue},
  pdfcreator={LaTeX via pandoc}}

\title{Informative Distance-Based Priors for Correlation Matrices
Centred on a Target Reference}
\author[1]{Anna
Freni-Sterrantino\thanks{Corresponding author: \href{mailto:afrenisterrantino@turing.ac.uk}{afrenisterrantino@turing.ac.uk}}}
\author[2]{Janet van Niekerk}
\author[3]{Elias Teixeira Krainski}
\author[4]{Denis Rustand}
\author[3,5]{Haziq Jamil}
\author[3]{Håvard Rue}
\affil[1]{The Alan Turing Institute, London, United Kingdom}
\affil[2]{Department of Statistics, University of
Pretoria, Pretoria, South Africa}
\affil[3]{Computer, Electrical and Mathematical Sciences and Engineering
(CEMSE) Division, King Abdullah University of Science and
Technology, Thuwal, Kingdom of Saudi Arabia}
\affil[4]{Biostatistics team - Bordeaux Population Health Research
Center, National Institute of Health and Medical Research (Inserm UMR
1219), Bordeaux, France}
\affil[5]{Mathematical Sciences, Faculty of Science, Universiti Brunei
Darussalam, Bandar Seri Begawan, Brunei}
\date{}
\begin{document}
\maketitle
\begin{abstract}
Specifying a prior over the space of correlation matrices is a persistent challenge in Bayesian analysis. The space is a curved manifold whose dimension grows quadratically with the number of variables, making substantive prior beliefs difficult to encode.\\
We propose a distance-based prior that assigns mass decaying exponentially in the Fisher arc-length distance from a user-specified reference correlation matrix, enabling shrinkage toward any target correlation structure rather than being confined to the identity matrix.
Formally, this is constructed as a Penalised Complexity prior, but its interpretation shifts accordingly: unless the chosen target represents a structurally simpler state, the shrinkage penalises deviation rather than complexity in the usual sense.
To accommodate conditional independence constraints, we introduce a parameterisation that constructs the correlation matrix via the Cholesky factor of the inverse correlation matrix with respect to a user-supplied graph, thereby reducing the number of free parameters from one per variable pair to one per graph edge.
The prior is proper for every positive value of its rate parameter, accommodates correlations of either sign under any graph structure, and reduces to a fully unstructured prior when the graph is complete.
A direct sampling algorithm is provided, enabling prior predictive checks and sensitivity analysis, implemented within the \texttt{graphpcor} package.
\end{abstract}
\par
\clearpage

\setstretch{2}

\section{Introduction}\label{sec-intro}

Correlation matrices are fundamental to multivariate Bayesian modelling, but specifying coherent priors over them remains a persistent challenge. They appear as parameters of multivariate Gaussian likelihoods, as components of hierarchical priors for random effects, and as structural objects in Gaussian graphical models.
They play a central role in meta-analysis, longitudinal modelling, network psychometrics, and spatial statistics, among many others.
Despite its ubiquity, specifying a prior over the space of \(p \times p\) correlation matrices \(\mathcal R_p\)---symmetric, positive definite, and with unit diagonal, a curved manifold of dimension \(p(p-1)/2\)---is substantially more demanding than specifying a prior over an unconstrained parameter.

Conjugate priors for the full covariance matrix have long been the standard choice. The Wishart and inverse-Wishart distributions \citep{wishart1928generalised} are widely used because they yield tractable posteriors under Gaussian likelihoods.
However, their fundamental shortcoming is the artificial coupling of marginal variances and correlations within a single hyperparameter structure.
Changing the scale matrix or degrees of freedom perturbs variances and correlations simultaneously, with no clean interpretation \citep{bekker2017wishart, tokuda2025visualizing}.
\citet{barnard2000modeling} proposed the \emph{separation strategy}, assigning independent priors to marginal standard deviations and the correlation matrix \(R\).
This decoupling is now standard, and it is also the viewpoint adopted throughout this paper.

Two broad approaches exist for specifying priors over \(R\).
Element-wise priors assign marginal distributions to the individual correlations \(\rho_{ij}\), for instance through the beta distribution or a normal prior on the Fisher \(z\)-transform, but these do not constrain the joint matrix to be positive definite and can place mass on infeasible matrices outside \(\mathcal{R}_p\) \citep{daniels1999nonconjugate}.
Joint priors for \(R\), on the other hand, avoid this by construction, typically through an unconstrained parameterisation that maps to \(\mathcal{R}_p\).
The most widely adopted is the LKJ distribution \citep{lewandowski2009generating}, which assigns density proportional to \(|R|^{\eta-1}\) and readily available in Stan \citep{carpenter2017stan} and PyMC \citep{pymc2023}.

The LKJ prior is governed by a single shape parameter \(\eta\): for \(\eta = 1\) it is uniform over \(\mathcal{R}_p\), and for \(\eta > 1\) it concentrates mass around the identity matrix \(I_p\).
Two drawbacks are noteworthy for applied work.
First, all correlations receive identically distributed marginals, so there is no mechanism for encoding that certain pairs of variables are expected to be more strongly correlated than others.
Second, the only attainable centre of contraction is the identity \(I_p\), since the single parameter \(\eta\) tunes the strength of shrinkage toward independence but cannot recentre the prior on a substantively motivated based correlation \(R \neq I_p\).
A related effect, that for fixed \(\eta\) the concentration toward \(I_p\) strengthens with the dimension \(p\) \citep{tokuda2025visualizing}, is an inherent consequence of the geometry of \(\mathcal{R}_p\) rather than a defect, but it does mean that \(\eta\) alone cannot express a dimension-stable prior belief about distance from a target.

Addressing these limitations, \citet{freni2025graphical} introduced a framework for correlation matrix priors that reduces the parameter count from \(p(p-1)/2\) to the number of latent variables, adopting a shared-latent-variable approach in which the correlation matrix \(R\) is induced by a parent-children relation on a tree structure.
Within this graphical framework, a Penalised Complexity (PC) prior \citep{simpson2017penalising} assigns mass that decays exponentially in the Kullback Leibler divergence (KLD) from a sequence of progressively simpler models.
The approach represents genuine progress, but limitations remain.
The latent Cholesky parameterisation underpinning the construction confines each correlation to a fixed sign orthant, so that a configuration which begins positive stays positive and the data cannot move a correlation across zero.
More fundamentally, and as with the LKJ prior, the contraction is ultimately toward the identity matrix.

In this work, we propose a prior for \(R\) that resolves these limitations.
The approach rests on two ideas.
First, we adopt the PC prior construction of \citet{simpson2017penalising} directly on the space of correlation matrices, where distance from a user-specified \emph{reference} correlation matrix \(R_0\) is measured via the square-root Kullback-Leibler divergence  between the corresponding Gaussian distributions, and the prior assigns mass that decays exponentially in this distance.
A local quadratic approximation to the KLD yields the Fisher metric on \(\mathcal{R}_p\); its matrix square root whitens the curved geometry of \(\mathcal{R}_p\) into a Euclidean space amenable to this distance-based construction.
Second, we parameterise \(R\) through the Cholesky factor of the precision matrix \(Q = R^{-1}\).
For a graph \(\mathcal{G} = (\mathcal{V}, \mathcal{E})\) that encodes the desired conditional independence structure, this reduces the number of free parameters from \(p(p-1)/2\) to \(|\mathcal{E}|\).
The remaining Cholesky entries---the \emph{fill-in} elements---are determined analytically from the free parameters, so the bijection \(\theta \mapsto R(\theta)\) is explicit and differentiable throughout.
This construction adapts triangular completion, an established technique used for G-Wishart computation in Gaussian graphical models \citep{roverato2002hyper, atay2005monte}.

The prior has three ingredients specified by the user:
(a) a reference (or target) correlation matrix \(R_0 \in \mathcal{R}_p\); (b) a graph \(\mathcal{G}\) encoding conditional independence, with the complete graph recovering the dense, fully unstructured prior; (c) a scalar rate \(\lambda > 0\) governing the strength of contraction toward \(R_0\).
Both positive and negative correlations are accommodated under any graph structure, the prior is proper for all \(\lambda > 0\), and the identity \(R_0 = I_p\) recovers a natural default that shares the shrinkage-toward-independence orientation of the LKJ prior whilst retaining graph-structured sparsity.

The paper is organised as follows.
Section~\ref{sec-preliminaries} reviews the required background on multivariate Gaussian and graphical models, the fill-in theory for Cholesky factors of sparse precision matrices, information geometry, and the PC prior framework.
Section~\ref{sec-prior} derives the prior in full generality, establishes its properties, and illustrates its behaviour.
Section~\ref{sec-sparse} develops the graphical Cholesky parameterisation that imposes the conditional independence structure.
Section~\ref{sec-data} applies the prior to a real dataset, and Section~\ref{sec-conclusion} concludes.

\section{Preliminaries}\label{sec-preliminaries}

We present the necessary background here, including Gaussian graphical models, Cholesky fill-in theory, information geometry, and Penalised Complexity priors.

\subsection{Multivariate Gaussian and Graphical Models}\label{sec-mvn}

Throughout the paper we work with a \(p\)-dimensional random vector \(x = (x_1, \ldots, x_p)^{\top}\) following the multivariate Gaussian distribution \(\operatorname{N}_p(\mu, \Sigma)\), with mean \(\mu \in \mathbb{R}^p\) and positive definite covariance matrix \(\Sigma \in \mathbb{R}^{p \times p}\).
Following the \emph{separation strategy} of \citet{barnard2000modeling}, the covariance admits the decomposition
\begin{equation}\protect\phantomsection\label{eq-separation}{
\Sigma = D R D, \qquad D = \operatorname{diag}(\sigma_1, \ldots, \sigma_p), \qquad R_{ij} = \Sigma_{ij} / (\sigma_i \sigma_j),
}\end{equation}
which disentangles the marginal standard deviations in \(D\) from the correlation matrix \(R\), and permits the two components to be modelled independently a priori.
While a fully Bayesian framework estimates \(\mu\) and \(D\) jointly with \(R\) to properly propagate uncertainty, the specification of their priors and their subsequent computation are standard and can be handled by routine means.
We therefore focus exclusively on the correlation matrix.

The inverse correlation matrix \(Q := R^{-1}\) provides a natural representation for encoding conditional independence among the components of \(x\); we refer to it simply as the precision throughout.
Whereas \(R\) records marginal correlations, \(Q\) records conditional ones.
For \(i \neq j\), its off-diagonal entries determine the partial correlation between \(x_i\) and \(x_j\) given the remaining variables,
\begin{equation}\protect\phantomsection\label{eq-partialcorr}{
\operatorname{Corr}(x_i, x_j \mid x_{-ij}) = - Q_{ij} \big/ \sqrt{Q_{ii} Q_{jj}},
}\end{equation}
so that the off-diagonal sparsity pattern of \(Q\) coincides with the conditional independence structure of \(x\),
\begin{equation}\protect\phantomsection\label{eq-hammersley}{
Q_{ij} = 0 \quad \iff \quad x_i \perp x_j \,\,\big|\,\, x_{-ij}.
}\end{equation}
Zeroes in \(Q\) as a modelling device originate with \citet{dempster1972covariance}, while their reading as conditional independence---the Gaussian case of the Hammersley-Clifford theorem---was established by \citet{speed1986gaussian} and \citet{lauritzen1996graphical}; see also \citet{drton2017structure}.

We encode the conditional independence pattern of \(x\) through an undirected graph \(\mathcal{G} = (\mathcal{V}, \mathcal{E})\) on \(\mathcal{V} = \{1, \ldots, p\}\), with edge set \(\mathcal{E} \subseteq \{\, (i, j) : 1 \le i < j \le p \,\}\).
The random vector \(x\) is a \emph{Gaussian Markov random field} (GMRF) with respect to \(\mathcal{G}\) if \(Q_{ij} = 0\) whenever \((i, j) \notin \mathcal{E}\); see \citet{rue2005gaussian} for a comprehensive treatment.
The graph \(\mathcal{G}\) thus specifies a prescribed sparsity pattern for \(Q\), and the number of free off-diagonal parameters equals the number of edges,
\[
n_Q = |\mathcal{E}| \le \tbinom{p}{2},
\]
with equality recovering the dense (fully connected) case.
A sparse graph \(\mathcal{G}\) aids interpretation, since partial correlations along non-edges are zero by construction.
To work with \(Q\), we parameterise it through its Cholesky factor \(L\), an unconstrained route to positive definite precision matrices; such Cholesky parameterisations are a common device in correlation modelling, the LKJ prior of \citet{lewandowski2009generating} being a prominent example.

\subsection{Cholesky Fill-in}\label{sec-fillin}

Adopting the notation of the previous subsection, we call a symmetric matrix \(Q \in \mathbb{R}^{p \times p}\) \textbf{\(\mathcal{G}\)-sparse} if \(Q_{ij} = 0\) for every \(i \neq j\) with \(\{i,j\} \notin \mathcal{E}\), and write \(\mathcal{S}_{++}^{p}\) for the cone of \(p \times p\) symmetric positive definite matrices and \(\mathcal{S}_{++}^{p}(\mathcal{G}) := \{Q \in \mathcal{S}_{++}^{p} : Q \text{ is } \mathcal{G}\text{-sparse}\}\).
For \(Q \in \mathcal{S}_{++}^{p}\), let \(L = L(Q)\) denote the unique lower-triangular matrix with positive diagonal satisfying \(Q = LL^{\!\top}\).
The entries of \(L\) can be determined column by column for \(j = 1, \ldots, p\) via
\begin{equation}\protect\phantomsection\label{eq-prelim-chol}{
L_{jj} = \bigg( Q_{jj} - \sum_{k < j} L_{jk}^{2} \bigg)^{1/2}, \qquad
L_{ij} = \frac{Q_{ij} - \sum_{k < j} L_{ik}\,L_{jk}}{L_{jj}}, \quad i > j.
}\end{equation}

\begin{figure}[!htbp]

\centering{

\includegraphics[width=4.9875in,height=\textheight,keepaspectratio]{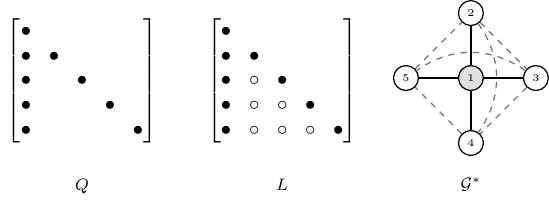}

}

\caption{\label{fig-three-panels}Star graph \(\mathcal{G}\) with \(p=5\) nodes and the hub at vertex 1. Shown are the lower-triangular sparsity patterns for \(Q\) (left) and its Cholesky factor \(L\) (right); \(\bullet\) denotes a structurally non-zero entry and a blank denotes a structural zero. In \(L\), \(\circ\) denotes generically non-zero fill-in entries at non-edge positions. The filled graph \(\mathcal{G}^{*}\) (right) shows solid edges (\(\mathcal{E}\)) matching the non-zero entries of \(Q\), and dashed edges (\(\mathcal{E}^{*} \setminus \mathcal{E}\)) show the fill-in positions. With the hub first, all \(\binom{4}{2}=6\) leaf pairs become fill, leaving \(\mathcal{I}_{\mathrm{zero}}=\emptyset\).}

\end{figure}

When \(Q\) is \(\mathcal{G}\)-sparse, its Cholesky factor \(L(Q)\) does not in general inherit the same sparsity pattern.
Figure~\ref{fig-three-panels} illustrates this, showing that a star-graph sparsity pattern on \(Q\) yields a fully dense \(L\).
The locations of such entries are governed by the \emph{filled graph} of \citet[Lemma 4]{rose1976algorithmic}, recalled below.

\begin{definition}[Filled graph]\protect\hypertarget{def-prelim-filled}{}\label{def-prelim-filled}

The filled graph of \(\mathcal{G}\) is the undirected graph \(\mathcal{G}^{*} = (\mathcal{V}, \mathcal{E}^{*})\) defined by \(\{i, j\} \in \mathcal{E}^{*}\) if and only if there exists a path \(p_{ij}=[i \equiv v_0, v_1, \ldots, v_t \equiv j]\) in \(\mathcal{G}\) such that
\[
v_k < \min(i, j) \text{ for } 0 < k < t.
\]
Length-one paths \((t = 1)\) contain no interior vertex and are admissible, so \(\mathcal{E} \subseteq \mathcal{E}^{*}\).

\end{definition}

\begin{remark}[Separation and conditional independence]
The filled graph has an equivalent reading through the Markov structure of \(\mathcal{G}\).
For \(i < j\), let \(\mathcal{F}(i, j) = \{i+1, \ldots, j-1, j+1, \ldots, p\}\) denote the \emph{future} of \(i\) other than \(j\).
Then \citep[Sec 2.4]{rue2005gaussian}
\[
\{i, j\} \in \mathcal{E}^{*}
\ \iff \
\mathcal{F}(i, j) \text{ does not separate } i \text{ and } j \text{ in } \mathcal{G}
\ \iff\
x_i \not\perp x_j \mid x_{\mathcal{F}(i, j)} .
\]
A fill-in edge thus marks a pair \((i, j)\) that remains conditionally dependent given the future variables, so the structural zeroes of \(L\) are dictated by the conditional independence structure of \(\mathcal{G}\) alone, independently of the numerical values of \(Q\).
Note that the conditioning on the smaller set \(\mathcal{F}(i, j)\) rather than all other variables \(x_{-ij}\) as in (\ref{eq-hammersley}) separates fewer pairs, which is the source of the fill-in \(\mathcal{E} \subseteq \mathcal{E}^{*}\).
\end{remark}

The edges of the filled graph \(\mathcal{G}^*\) completely characterise the sub-diagonal sparsity pattern of \(L\), whereby an entry \(L_{ij}\) (with \(i>j\)) is structurally non-zero if and only if \(\{i,j\} \in \mathcal{E}^*\).
Formally, the filled graph induces a partition of the strictly lower-triangular index set \(\mathcal{I} := \{(i,j) : 1 \leq j < i \leq p\}\) into the three disjoint classes

\begin{equation}\protect\phantomsection\label{eq-prelim-partition}{
\begin{gathered}
\mathcal{I}_{\mathrm{edge}} = \{(i,j) \in \mathcal{I} : \{i,j\} \in \mathcal{E}\}, \qquad
\mathcal{I}_{\mathrm{fill}} = \{(i,j) \in \mathcal{I} : \{i,j\} \in \mathcal{E}^{*} \setminus \mathcal{E}\}, \\[0.5em]
\mathcal{I}_{\mathrm{zero}} = \{(i,j) \in \mathcal{I} : \{i,j\} \not\in \mathcal{E}^{*} \},
\end{gathered}
}\end{equation}

of cardinalities \(n_Q\), \(|\mathcal{E}^{*}| - n_Q\), and \(\binom{p}{2} - |\mathcal{E}^{*}|\), respectively.
With this taxonomy established, the following lemma records how each class manifests in the Cholesky factor.
The result is classical, restating standard sparse-Cholesky fill-in theory \citep{rose1976algorithmic, davis2006direct, rue2005gaussian} in the notation of this paper.
We include it for completeness and to support the parameterisation of Section~\ref{sec-sparse}.

\begin{lemma}[Three-class structure of \(L\)]\protect\hypertarget{lem-prelim-three-class}{}\label{lem-prelim-three-class}

Let \(Q \in \mathcal{S}_{++}^{p}(\mathcal{G})\) and \(L = L(Q)\). For every \((i, j) \in \mathcal{I}\),

\begin{enumerate}
\def\labelenumi{(\roman{enumi})}
\item
  if \((i, j) \in \mathcal{I}_{\mathrm{zero}}\), then \(L_{ij} = 0\);
\item
  if \((i, j) \in \mathcal{I}_{\mathrm{fill}}\), then \(L_{ij}\) is a deterministic function of the preceding edge entries \(\{Q_{rs} : (r,s) \in \mathcal{I}_{\mathrm{edge}},\, s < j\}\) and the diagonal entries \(\{Q_{kk} : k < j\}\).
\item
  the entries \(\{L_{ij} : (i, j) \in \mathcal{I}_{\mathrm{edge}}\}\) together with the positive fixed diagonal \((L_{11}, \ldots, L_{pp})\) uniquely determine \(L\) via (\ref{eq-prelim-chol}).
\end{enumerate}

\end{lemma}

The proof, given in the Appendix, is a straightforward induction on the column index \(j\).
The key structural result is that entries at \(\mathcal{I}_{\mathrm{zero}}\) positions vanish identically for every \(Q \in \mathcal{S}_{++}^{p}(\mathcal{G})\), whereas entries at \(\mathcal{I}_{\mathrm{fill}}\) positions are, in general, non-zero.
More precisely, a fill entry \(L_{ij}\) with \((i,j) \in \mathcal{I}_{\mathrm{fill}}\) is a nontrivial rational function of the edge and diagonal entries, so it vanishes only when those values conspire to cancel exactly---what \citet{rose1976algorithmic} term ``lucky cancellation''.

We remark that the statistical literature had independently derived these same mathematical facts.
Specifically, \citet{wermuth1980linear} established the fill-free factorisation of decomposable patterns using a perfect ordering.
This construction was later extended to arbitrary graphs by \citet[App. A]{roverato2002hyper} through the completion of incomplete triangular matrices.
Finally, \citet[Prop. 2]{atay2005monte} restated this completion process as an explicit recursion.
The latter two sources also provide a constructive algorithm for computing the fill-in entries, which we adapt in Section~\ref{sec-sparse} to parameterise the correlation matrix sparsely.

\begin{figure}[!htbp]

\centering{

\includegraphics[width=4.9875in,height=\textheight,keepaspectratio]{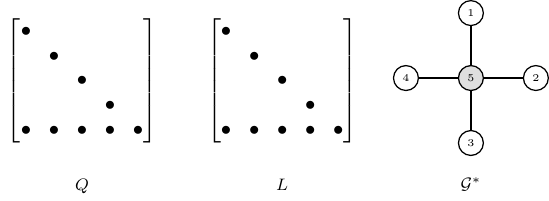}

}

\caption{\label{fig-three-panels-reordered}Same star graph as Figure~\ref{fig-three-panels} but with the hub relabelled as vertex 5 (hub last).
The lower-triangular sparsity patterns of \(Q\) and \(L\) are now identical.
No fill entries arise, and the filled graph \(\mathcal{G}^{*} = \mathcal{G}\) carries no dashed edges.}

\end{figure}

Finally, the partition of \(\mathcal I\) in (\ref{eq-prelim-partition}) and the filled graph \(\mathcal{G}^{*}\) depend on the labelling of \(\mathcal{V}\), not merely on the abstract graph \(\mathcal{G}\).
In the star example, placing the hub at vertex 5 instead of at 1 yields \(\mathcal{G}^{*} = \mathcal{G}\) with \(|\mathcal{I}_{\mathrm{fill}}| = 0\), and the Cholesky factor inherits the exact sparsity of \(Q\) as illustrated in Figure~\ref{fig-three-panels-reordered}.
Finding the permutation \(\pi\) that minimises \(|\mathcal{E}^{*}_{\pi}|\) is NP-hard \citep{yannakakis1981computing}.
Practical heuristics such as bandwidth reduction, nested dissection, and minimum-degree are surveyed in \citet[Ch. 2]{rue2005gaussian} and \citet[Ch. 7]{davis2006direct}.
A graph admits a fill-free ordering (\(\mathcal{E}^{*}_{\pi} = \mathcal{E}\) for some \(\pi\)) if and only if it is chordal \citep{rose1976algorithmic, wermuth1980linear, lauritzen1996graphical}.

\subsection{Information Geometry}\label{sec-info-geom}

A guiding principle in our construction is that the space of probability distributions, rather than the space of parameters, is the object on which discrepancy should be measured.
We summarise the minimal facts required from information geometry.
Standard references are the comprehensive treatment of \citet{amari2016information} and the classical differential-geometric background of \citet{boothby2003introduction} and \citet{murray1993differential}.

Let \(\mathcal{P} = \{ \pi_\theta : \theta \in \Theta \}\) be a regular parametric family of densities on a sample space \(\mathcal{X}\), with open parameter set \(\Theta \subseteq \mathbb{R}^d\) and suitable smoothness and identifiability conditions so that \(\mathcal{P}\) carries the structure of a \(d\)-dimensional smooth manifold with global coordinates \(\theta = (\theta_1, \ldots, \theta_d)\).
The statistically natural Riemannian metric on \(\mathcal{P}\) is given by the Fisher information matrix
\begin{equation}\protect\phantomsection\label{eq-fisher}{
I(\theta) = \mathbb{E}_{\pi_\theta} \left[ \nabla_\theta \log \pi_\theta(X) \, \nabla_\theta \log \pi_\theta(X)^{\top} \right],
}\end{equation}
whose entries \(g_{ij}(\theta) = [I(\theta)]_{ij}\) define the metric tensor on \(\mathcal{P}\).
The squared infinitesimal distance between two neighbouring distributions \(\pi_\theta\) and \(\pi_{\theta + d\theta}\) is
\begin{equation}\protect\phantomsection\label{eq-line-element}{
ds^2 = \sum_{i,j} g_{ij}(\theta)\, d\theta_i\, d\theta_j = d\theta^{\top} I(\theta)\, d\theta.
}\end{equation}
Crucially, because \(I(\theta)\) transforms covariantly under smooth reparameterisations, the line element in (\ref{eq-line-element}) is an intrinsic property of the family \(\mathcal{P}\) and does not depend on the particular coordinates chosen.

The Kullback-Leibler divergence from \(\pi_{\theta'}\) to \(\pi_\theta\),
\[
\operatorname{D}_{\mathrm{KL}}(\pi_\theta \,\Vert\, \pi_{\theta'}) = \int \pi_\theta(x) \log \frac{\pi_\theta(x)}{\pi_{\theta'}(x)}\, dx,
\]
provides the information-theoretic yardstick for comparing distributions.
Although not a metric (it is neither symmetric nor does it satisfy the triangle inequality), it is locally quadratic and, under mild regularity, agrees with the Fisher metric to leading order.
Expand \(f(d\theta) := \operatorname{D}_{\mathrm{KL}}(\pi_{\theta + d\theta} \,\Vert\, \pi_\theta)\) about \(d\theta = 0\), holding the reference \(\pi_\theta\) fixed in the second argument as in the prior construction of the sequel.
The linear term vanishes in the expansion because the expected score is zero, and the quadratic coefficient is precisely the Fisher information:
\begin{equation}\protect\phantomsection\label{eq-kld-taylor}{
\operatorname{D}_{\mathrm{KL}}(\pi_{\theta + d\theta} \,\Vert\, \pi_\theta) = \tfrac{1}{2}  d\theta^{\top} I(\theta)\, d\theta + o(\| d\theta \|^2).
}\end{equation}
Thus, to second order, KLD is half the squared Fisher arc length in (\ref{eq-line-element}).

The connection is not merely asymptotic.
The KLD is the canonical measure of statistical discrepancy; it quantifies the information lost when \(\pi_{\theta'}\) is used in place of \(\pi_\theta\).
The Fisher metric can equivalently be \emph{derived} from KL by asking: what Riemannian metric on \(\mathcal{P}\) is consistent with this discrepancy in the limit of nearby distributions?
Equation~\ref{eq-kld-taylor} shows the answer is unique, and it is \(I(\theta)\).
The two objects are therefore not merely analogous but canonically identified, and the Fisher metric is the infinitesimal geometry that KLD induces on \(\mathcal{P}\).

\subsection{Penalised Complexity Priors}\label{penalised-complexity-priors}

The Penalised Complexity (PC) prior framework of \citet{simpson2017penalising} specifies priors by penalising the \emph{distance} of a model from a designated base model, rather than directly on the parameters.
Given the identification between KLD and the Fisher metric in (\ref{eq-line-element}) and (\ref{eq-kld-taylor}), a natural choice of distance between \(\pi_\theta\) and a base distribution \(\pi_{\theta_0}\) is
\begin{equation}\protect\phantomsection\label{eq-pc-distance}{
d(\theta) = \sqrt{2 \operatorname{D}_{\mathrm{KL}}(\pi_\theta \,\Vert\, \pi_{\theta_0})\,} \approx \sqrt{(\theta - \theta_0)^{\top} I(\theta_0) (\theta - \theta_0)},
}\end{equation}
which has units of Fisher arc length and reduces, for \(\theta\) near \(\theta_0\), to the Riemannian distance induced by \(I(\theta_0)\).

Given this distance, the construction follows \citet{simpson2017penalising}.
By Occam's razor, the prior prefers the simpler base model in the absence of evidence to the contrary, penalising deviation so that its density decreases with \(d(\theta)\).
This penalisation is applied at a constant rate on the distance scale---the natural default when \(d\) carries no interpretation beyond distance itself---which singles out the exponential prior on \(d\),
\begin{equation}\protect\phantomsection\label{eq-pc-rate}{
\pi_d(d) = \lambda \exp(-\lambda d), \qquad d \ge 0,
}\end{equation}
where \(\lambda > 0\) is typically elicited through a tail probability statement \(\Pr\{ Q(\theta) > U \} = \alpha\) for some interpretable transformation \(Q\) of \(\theta\).
This last step gives the user explicit, interpretable control of the shrinkage strength.

When \(\theta\) is scalar and \(d\) one-to-one, a change of variables yields the prior on \(\theta\),
\begin{equation}\protect\phantomsection\label{eq-pc-prior}{
\pi(\theta) = \lambda \exp\!\big( -\lambda\, d(\theta) \big) \big| d'(\theta) \big|,
}\end{equation}
with \(d'(\theta)\) the ordinary derivative of \(d\).
For the multivariate \(\theta\) of interest here the map is many-to-one---an entire level set \(\{\theta : d(\theta) = r\}\) shares one distance---so the prior must also distribute mass across level sets, which we address in Section~\ref{sec-prior}.
The resulting prior is proper whenever \(d(\theta) \to \infty\) on the boundary of \(\Theta\), invariant to reparameterisation, and modal at \(\theta_0\).
These qualities make this construction a natural vehicle for the informative correlation priors developed next.

\section{An Informative Distance-Based Prior for Correlation Matrices: The General Case}\label{sec-prior}

We now apply the machinery of Section~\ref{sec-preliminaries} to construct a prior on the space of \(p \times p\) positive definite correlation matrices,
\[
\mathcal{R}_p := \big\{ R \in \mathbb{R}^{p \times p} : R_{ii} = 1, R \succ 0 \big\}.
\]
The set \(\mathcal{R}_p\) is a bounded, smooth submanifold of the SPD cone \(\mathcal{S}_{++}^{p}\) of dimension \(p(p-1)/2\), cut out by the \(p\) unit-diagonal constraints.
Any prior on \(\mathcal{R}_p\) must respect these constraints, which is nontrivial because \(\mathcal{R}_p\) is not a convex set.
We dissolve these constraints by working instead with an unconstrained reparameterisation, a smooth bijection
\begin{equation}\protect\phantomsection\label{eq-rtheta}{
R : \mathbb{R}^m \to \mathcal{R}_p, \qquad \theta \mapsto R(\theta),
}\end{equation}
with smooth inverse on \(\mathcal{R}_p\) and \(m \le p(p-1)/2\), so that any prior on the unrestricted \(\theta \in \mathbb{R}^m\) induces a proper, positive-definite-respecting prior on \(R\).

The map we adopt is the Cholesky factor of the precision \(Q = R^{-1}\).
Writing \(Q = L L^{\top}\) as in Section~\ref{sec-fillin}, the strictly lower-triangular real-valued entries of \(L\) serve as the free parameters; they give the so-called \emph{dense case}, with \(m = p(p-1)/2\).
Factoring the precision rather than \(R\) itself is what lets us exploit graphical structure, since the conditional independence pattern lives in \(Q\).
When a graph \(\mathcal{G}\) prescribes this structure, only the entries of \(L\) tied to its edges remain free---the fill-in entries determined and the rest zero, by Lemma~\ref{lem-prelim-three-class}---so \(m\) reduces to the edge count \(|\mathcal{E}|\), as developed in Section~\ref{sec-sparse}.
The LKJ construction of \citet{lewandowski2009generating} instead Cholesky-factors \(R\) directly, where no such sparsity is exposed.
In the dense case the choice is immaterial, however, and the derivation below applies equally to the LKJ parameterisation, or to any other smooth bijection of the form (\ref{eq-rtheta}).

Let \(R_0 := R(\theta_0)\) denote a user-specified \emph{reference correlation matrix} encoding prior structural beliefs.
Following the separation strategy, we gauge the departure of a candidate \(R(\theta)\) from \(R_0\) by the Kullback-Leibler divergence (KLD) between the corresponding centred Gaussian models for \(x\) introduced in Section~\ref{sec-mvn}, which share a common mean and marginal scales.
The shared parameters cancel, leaving a divergence that depends on the correlation structure alone,
\begin{equation}\protect\phantomsection\label{eq-kld-gauss}{
\operatorname{D}_{\mathrm{KL}}\!\Big( \operatorname{N}\big(0, R(\theta)\big) \,\Big\Vert\, \operatorname{N}(0, R_0) \Big)
=
\frac{1}{2}\Big[
  \operatorname{tr}\!\big(R_0^{-1} R(\theta)\big)
  - p
  - \log |R(\theta)|
  + \log |R_0|
\Big].
}\end{equation}
This is non-negative, is finite for all \(R(\theta), R_0 \in \mathcal{R}_p\), and vanishes if and only if \(R(\theta) = R_0\).
The PC distance (\ref{eq-pc-distance}) then takes the form \(d(\theta) = \sqrt{2\,\operatorname{D}_{\mathrm{KL}}(\operatorname{N}(0, R(\theta)) \,\Vert\, \operatorname{N}(0, R_0))}\).
By the local quadratic identity (\ref{eq-kld-taylor}) established in Section~\ref{sec-info-geom}, near \(\theta_0\) the KLD is approximated by a quadratic form whose matrix is the Fisher information,
\begin{equation}\protect\phantomsection\label{eq-hessian}{
H(\theta_0) := \nabla^2_\theta \operatorname{D}_{\mathrm{KL}} \Big( \operatorname{N}\big(0, R(\theta)\big) \,\Big\Vert\, \operatorname{N}(0, R_0) \Big)\Bigg|_{\theta = \theta_0},
}\end{equation}
computed once, offline, at the reference model \(R_0\).

In the case when \(R_0 = I_p\), the trace term in (\ref{eq-kld-gauss}) satisfies \(\operatorname{tr}(R(\theta)) = p\) for any correlation matrix, and \(\log|I_p| = 0\), so the KLD reduces to \(\operatorname{D}_{\mathrm{KL}} = -\tfrac{1}{2}\log|R(\theta)|\).
Departure from independence is thus measured by the log-determinant of \(R\) alone---a classical scalar summary of multivariate association, non-negative by Hadamard's inequality (\(|R| \le 1\) for any correlation matrix, with equality if and only if \(R = I_p\)).
Correspondingly, the Hessian reduces to
\[
H(\theta_0)_{jk} = \frac{1}{2}\operatorname{tr} \left(\frac{\partial R(\theta)}{\partial\theta_j}\bigg|_{\theta = \theta_0} \frac{\partial R(\theta)}{\partial\theta_k}\bigg|_{\theta = \theta_0}\right),
\]
the Frobenius inner product of the Jacobian matrices at the reference point.

Define the \emph{whitened parameter}
\begin{equation}\protect\phantomsection\label{eq-whitening}{
\xi := H(\theta_0)^{1/2} (\theta - \theta_0) \in \mathbb{R}^m,
}\end{equation}
where \(H(\theta_0)^{1/2}\) is the symmetric matrix square root.
By (\ref{eq-kld-taylor}), \(\operatorname{D}_{\mathrm{KL}} \approx \tfrac{1}{2}\|\xi\|^2\) in a neighbourhood of \(\theta_0\), where \(\|\xi\|\) is the Mahalanobis distance between \(\theta\) and \(\theta_0\) under the Fisher metric \(H(\theta_0)\).
Conesequently, the Euclidean norm
\begin{equation}\protect\phantomsection\label{eq-r}{
r := \|\xi\| \approx d(\theta)
}\end{equation}
is the natural scalar measure of departure from the reference model in whitened coordinates.
In this coordinate system the local Fisher geometry is isotropic, i.e.~all directions of departure from \(\theta_0\) are penalised equally, at a rate determined solely by \(r\).
By absorbing the arbitrary scales and curvatures of the original parameterisation, this whitening ensures that a spherically symmetric prior on \(\xi\) is geometrically well-posed.

To this end, write \(\xi = r u\) in hyperspherical coordinates, with \(r \ge 0\) scalar, and \(u \in \mathbb{S}^{m-1}\), the unit sphere in \(\mathbb{R}^m\).
Following the same constant-rate (exponential) penalisation on the distance scale, and with the absence of any directional preference, we assign
\begin{equation}\protect\phantomsection\label{eq-prior-spherical}{
r \mid \lambda \sim \operatorname{Exp}(\lambda), \qquad {u} \mid r \sim \operatorname{Uniform}(\mathbb{S}^{m-1}),
}\end{equation}
independently, where \(\lambda > 0\) governs the concentration around the reference model.
The surface area of \(\mathbb{S}^{m-1}\) is \(2\pi^{m/2}\big/\,\Gamma(m/2)\) and the hyperspherical Jacobian contributes \(r^{m-1}\), giving the joint density of \(\xi\) in Cartesian coordinates
tt\begin{equation}\protect\phantomsection\label{eq-prior-xi}{
\pi(\xi ; \lambda) = \frac{\Gamma(m/2)}{2\pi^{m/2}} \, \frac{\lambda e^{-\lambda r}}{r^{m-1}}, \qquad r = \|\xi\|.
}\end{equation}
For \(m = 1\), (\ref{eq-prior-xi}) reduces to the Laplace distribution \(\pi(\xi ; \lambda) = (\lambda/2)\exp(-\lambda|\xi|)\), recovering the univariate PC prior of \citet{simpson2017penalising}.
Inverting (\ref{eq-whitening}), the prior on \(\xi\) pushes forward to the unconstrained parameter \(\theta\) through the linear Jacobian \(|\det H(\theta_0)|^{1/2}\):
\begin{equation}\protect\phantomsection\label{eq-prior-theta}{
\pi(\theta ; \theta_0, \lambda)
=
\frac{\Gamma(m/2)}{2\pi^{m/2}} \,
\frac{\lambda e^{-\lambda r}}{r^{m-1}} \,
\big|\det H(\theta_0)\big|^{1/2},
\qquad r = \big\|H(\theta_0)^{1/2}(\theta - \theta_0)\big\|.
}\end{equation}

\subsection{Properties and Illustration}\label{properties-and-illustration}

For completeness, we record the elementary properties of the prior used in the sequel.
Each follows immediately from the radial-exponential construction or from standard PC-prior theory \citep{simpson2017penalising}.

\begin{proposition}[Properties of the informative correlation prior]\protect\hypertarget{prp-properties}{}\label{prp-properties}

Let \(\lambda > 0\), \(R_0 \in \mathcal{R}_p\), and let \(R(\theta)\) be any smooth bijection of the form (\ref{eq-rtheta}). Then the prior defined by (\ref{eq-prior-xi}) and (\ref{eq-prior-theta}) satisfies:

\begin{enumerate}
\def\labelenumi{(\roman{enumi})}
\item
  \emph{\textbf{Propriety}}. The density (\ref{eq-prior-xi}) integrates to one over \(\mathbb{R}^m\) for every \(\lambda > 0\) and \(m \ge 1\).
\item
  \emph{\textbf{Contraction to the reference model}}. The density (\ref{eq-prior-xi}) is a strictly decreasing function of \(r = \|\xi\|\). For \(m = 1\) it attains a finite maximum at \(\xi = 0\); for \(m \ge 2\) it is unbounded at the origin but remains integrable, and all level sets \(\{\xi : \pi(\xi ; \lambda) \ge c\}\) are balls centred at \(\theta_0\).
\item
  \emph{\textbf{Elicitation}}. Given \(R_0\), the prior is fully specified by the scalar \(\lambda\), which admits the tail-probability elicitation \(\Pr\{r > t\} = \alpha \iff \lambda = -\log(\alpha)/t\).
\end{enumerate}

\end{proposition}

Propriety (i) holds by construction, since (\ref{eq-prior-xi}) is the Cartesian density of independent \(r \sim \operatorname{Exp}(\lambda)\) and \(u \sim \operatorname{Uniform}(\mathbb{S}^{m-1})\).
The elicitation identity (iii) is the exponential tail \(\Pr\{r > t\} = e^{-\lambda t}\) of \(r\), and the contraction claims in (ii) are established in the Appendix.

Property (ii) deserves comment.
The unboundedness of (\ref{eq-prior-xi}) at \(\xi = 0\) when \(m \ge 2\) is not a pathology.
The density is defined against Lebesgue measure on \(\mathbb{R}^m\), on which every hypersphere of radius \(r\) has zero measure, so local integrability is preserved and posterior inference is unaffected.
What \emph{is} preserved universally is the geometric interpretation: for any \(\lambda > 0\) and any \(m \ge 1\), the prior places more mass on neighbourhoods of \(R_0\) of any prescribed Fisher radius than on annuli at larger radii, with the rate of contraction governed by \(\lambda\).
As a result, under any smooth bijection (\ref{eq-rtheta}) the induced prior concentrates on Fisher-geodesic neighbourhoods of \(R_0\).

The simplest instance is \(p = 2\), for which \(m = 1\) and the sole free parameter is the correlation \(\rho \in (-1, 1)\).
We derive the prior in closed form to illustrate the construction and its properties.

\begin{figure}[htbp]

\centering{

\includegraphics[width=1\linewidth,height=\textheight,keepaspectratio]{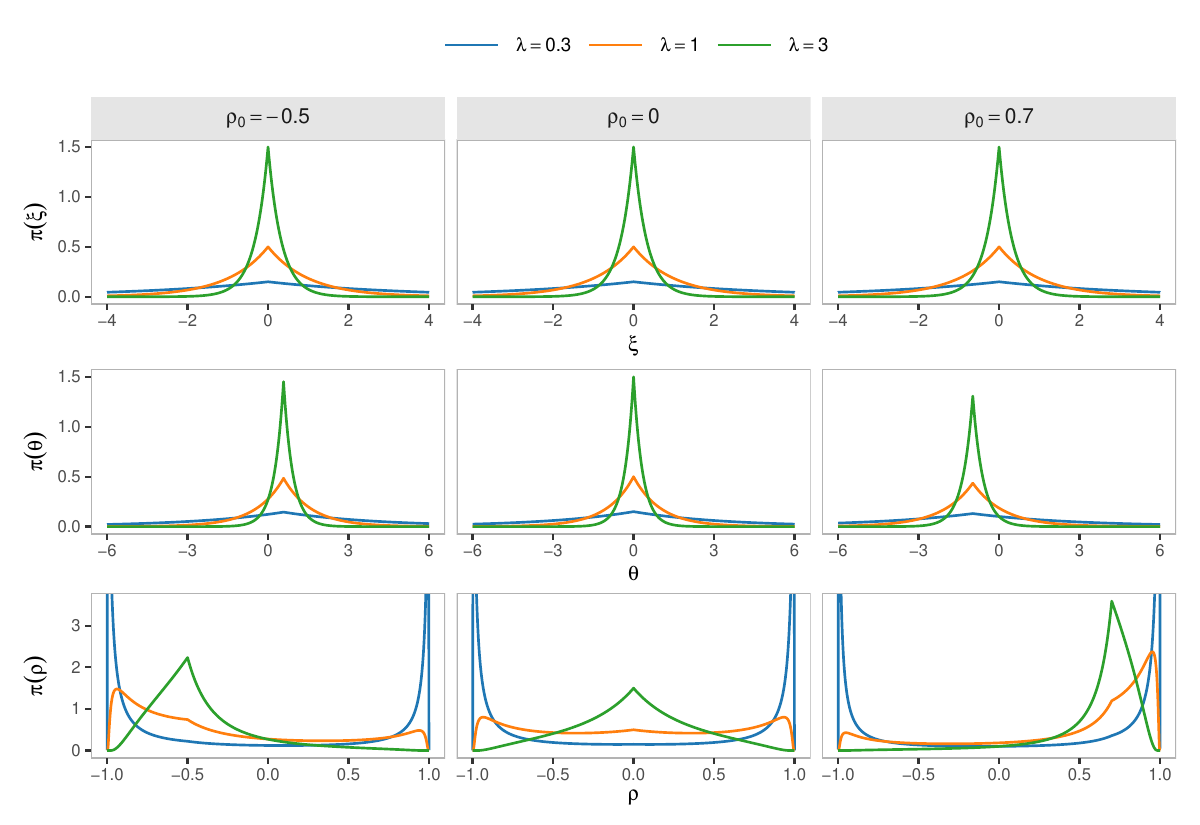}

}

\caption{\label{fig-uni-prior}Informative prior on the bivariate correlation under three parameterisations (rows) and three reference models (columns). In the whitened coordinate \(\xi\), the prior is an exact Laplace distribution centred at zero, identical across all choices of \(\rho_0\) (top row). The precision-Cholesky coordinate \(\theta\) is related to \(\xi\) by the affine whitening (\ref{eq-whitening}), so the prior on \(\theta\) is again Laplace, centred and scaled differently (middle row). In the native correlation \(\rho\), the nonlinearity of the parameterisation is pronounced (bottom row).}

\end{figure}

\begin{example}[The \(2 \times 2\) case]\protect\hypertarget{exm-prior-p2}{}\label{exm-prior-p2}

The correlation matrix and its precision are
\begin{equation}\protect\phantomsection\label{eq-p2-RQ}{
R = \begin{pmatrix} 1 & \rho \\ \rho & 1 \end{pmatrix},
\qquad
Q = R^{-1} = \frac{1}{1 - \rho^2}\begin{pmatrix} 1 & -\rho \\ -\rho & 1 \end{pmatrix},
}\end{equation}
and, introducing a single free parameter \(\theta \in \mathbb{R}\), we factorise the precision as \(Q = L L^{\top}\) with \(\theta\) placed in the lower-triangular factor below a fixed diagonal \((2, 1)\), a construction developed in the next section.
This gives
\begin{equation}\protect\phantomsection\label{eq-p2-LQ}{
L(\theta) = \begin{pmatrix} 2 & 0 \\ \theta & 1 \end{pmatrix},
\qquad
Q(\theta) = L(\theta)\,L(\theta)^{\top} = \begin{pmatrix} 4 & 2\theta \\ 2\theta & \theta^2 + 1 \end{pmatrix}.
}\end{equation}
Normalising \(V(\theta) = Q(\theta)^{-1}\) to unit diagonal returns a matrix of the form (\ref{eq-p2-RQ}), with the free parameter and correlation related by
\begin{equation}\protect\phantomsection\label{eq-p2-map}{
\rho(\theta) = -\frac{\theta}{\sqrt{1 + \theta^2}},
\qquad
\theta(\rho) = -\frac{\rho}{\sqrt{1 - \rho^2}},
\qquad
\left| \frac{\mathrm{d}\rho}{\mathrm{d}\theta} \right| = (1 + \theta^2)^{-3/2} = (1 - \rho^2)^{3/2}.
}\end{equation}
The off-diagonal \(Q_{12} = 2\theta\) carries the opposite sign to \(\rho\), and \(\theta\) ranges over all of \(\mathbb{R}\) as \(\rho\) traverses \((-1, 1)\).

Evaluating the KLD Hessian (\ref{eq-hessian}) at the reference point \(\theta_0 = \theta(\rho_0)\) gives the Fisher information for \(\theta\), which by the change of coordinates (\ref{eq-p2-map}) is the Fisher information \(I(\rho_0)\) for the correlation scaled by the squared Jacobian,
\begin{equation}\protect\phantomsection\label{eq-p2-H}{
H(\theta_0) = I(\rho_0)\,\bigg|\frac{\mathrm{d}\rho}{\mathrm{d}\theta}\bigg|^{2} =\frac{1 + \rho_0^2}{(1 - \rho_0^2)^2}\,(1 - \rho_0^2)^3 = 1 - \rho_0^4.
}\end{equation}
The whitened coordinate is therefore \(\xi = (1 - \rho_0^4)^{1/2}(\theta - \theta_0)\) and, by the \(m = 1\) reduction of (\ref{eq-prior-xi}), the prior on \(\xi\) is the Laplace law \(\tfrac{\lambda}{2}\exp(-\lambda|\xi|)\), symmetric and identical in shape across all \(\rho_0\).

Because the whitening (\ref{eq-whitening}) is affine, the prior is an exact Laplace in \(\theta\), centred at \(\theta_0\) with rate \(\lambda H(\theta_0)^{1/2}\).
The subsequent map to \(\rho\) is nonlinear, so the induced correlation prior is a \(\rho_0\)-dependent reshaping of this Laplace: it tightens at \(\rho_0\) for large \(\lambda\), spreads toward the boundaries \(\pm 1\) for small \(\lambda\), and is asymmetric whenever \(\rho_0 \neq 0\).
Figure~\ref{fig-uni-prior} shows the prior in all three coordinates across three reference models and a range of \(\lambda\).

\end{example}

\protect\phantomsection\label{cell-fig-uni-prior}

A further structural feature concerns the prior's behaviour at the boundary.
The reader will recognise the pushforward density \(\rho\) as an exponentially decaying Laplace kernel multiplied by a polynomially growing Jacobian.
As \(\rho \to \pm 1\), the unconstrained coordinate runs off to infinity, so the exponential decay overwhelms the polynomial growth and the density falls to zero at the boundary for every \(\lambda > 0\).
This reshaping is shown in the \(\theta\) and \(\rho\) rows of Figure~\ref{fig-uni-prior}.

It is worth noting that the canonical PC prior construction of \citet{simpson2017penalising} presumes a simpler base model, though the actual mathematics of the construction does not require this.
The base distribution may be any model in the parameter space, and the prior (\ref{eq-prior-theta}) simply penalises distance from the base \(R(\theta_0)\) at a constant rate (\ref{eq-pc-rate}) regardless of its structure.
Changing the base \(R(\theta_0)\) alters only the \emph{interpretation} of complexity.
In many settings the base model is naturally a simpler one---zero variance for an i.i.d. random effect, or zero or unit correlation for an AR(1) process---and Occam's razor applies perfectly.
For correlation matrices, no such structural simplification is at stake, and the natural choice of a base model is instead a user-specified reference correlation structure, which need not be ``simple'' in any structural sense.
Complexity is then regarded as a departure from that reference, and the prior is informative in the sense that it encodes a preference for models close to the reference as user-provided.

\section{An Informative Distance-Based Prior for Correlation Matrices: The Graph Case}\label{sec-sparse}

The  parameterisation of Section~\ref{sec-prior} leaves all \(\binom{p}{2}\) sub-diagonal entries of \(L\) free, none of them tied to the conditional independence structure of \(x\).
That structure is encoded in the sparsity pattern of the precision \(Q = R^{-1}\) (recall Equation~\ref{eq-hammersley}), and respecting it lets us parameterise \(R\) with fewer than \(\binom{p}{2}\) parameters.
Given a user-supplied graph \(\mathcal{G} = (\mathcal{V}, \mathcal{E})\) that, as in Section~\ref{sec-mvn}, prescribes the sparsity pattern of \(Q\), we construct a smooth forward map \(\theta \mapsto R(\theta)\) with \(m = n_Q\) free parameters, one per edge, whose image consists of correlation matrices with \(\mathcal{G}\)-sparse precision.
The sparse parameterisation is grounded in the three-class structure of Lemma~\ref{lem-prelim-three-class} and builds upon the triangular completion method from Gaussian graphical models \citep{roverato2002hyper, atay2005monte}.
By adapting this established technique to the correlation manifold, the unit-diagonal constraint fixes the Cholesky diagonal, leaving exactly one free parameter for each edge.
We call the resulting map \(\theta \mapsto R(\theta)\) the \textbf{Graphical Cholesky Parameterisation} (GCP), where \(L(\theta)\) is populated column by column as follows:

\begin{enumerate}
\def\labelenumi{\arabic{enumi}.}
\tightlist
\item
  \emph{Diagonal entries.} Fix \(L_{ii} = d_i > 0\) for \(i = 1, \ldots, p\), with \(d_i = p - i + 1\) as a suggested choice.
\item
  \emph{Edge entries.} For \((i,j) \in \mathcal{I}_{\mathrm{edge}}\), the corresponding entry \(L_{ij}\) is free; assign \(L_{ij}(\theta) = \theta_{k}\), where the index \(k \in \{1,\dots,n_Q\}\) increments sequentially as the edge positions are traversed in column-major order.
\item
  \emph{Non-edge entries.} For \((i,j) \not\in \mathcal{I}_{\mathrm{edge}}\), set
  \begin{equation}\protect\phantomsection\label{eq-gcp-fillin}{
  L_{ij}(\theta) = -\frac{1}{d_j} \sum_{k=1}^{j-1} L_{ik}(\theta)\, L_{jk}(\theta),
  }\end{equation}
  applied column by column so that each right-hand side involves only previously assigned entries.
\end{enumerate}

By Lemma~\ref{lem-prelim-three-class}, this recipe uniquely determines all of \(L(\theta)\) from \(\theta\) and \(d_1, \ldots, d_p\) alone, as entries at \(\mathcal{I}_{\mathrm{zero}}\) vanish identically, and entries at \(\mathcal{I}_{\mathrm{fill}}\) are generically non-zero rational functions of the \(n_Q\) free parameters, so the remaining non-edge positions carry no additional degrees of freedom.
In the limiting case of a complete graph, \(\mathcal{I}_{\mathrm{fill}} = \mathcal{I}_{\mathrm{zero}} = \emptyset\) and every sub-diagonal entry of \(L\) is free; the GCP then covers all of \(\mathcal{R}_p\), and working through \(Q = LL^\top\) or directly through \(R\) amounts to a choice of coordinates on the same space.
The correlation matrix is recovered by
\begin{equation}\protect\phantomsection\label{eq-forward-map}{
Q(\theta) = L(\theta) L(\theta)^{\!\top}, \quad V(\theta) = Q(\theta)^{-1}, \quad
R(\theta) = \operatorname{diag}(V(\theta))^{-1/2}\, V(\theta)\, \operatorname{diag}(V(\theta))^{-1/2},
}\end{equation}
where the final step normalises \(V(\theta)\) to unit diagonal.
No constraint on \(\theta\) is required, and the construction is well-defined on \(\mathbb{R}^{n_Q}\).

\begin{remark}[Choice of diagonal]
The forward map (\ref{eq-forward-map}) is invariant to the choice of positive diagonal values.
Scaling row \(i\) of \(L\) by any \(c_i > 0\) maps \(Q_{ij} \mapsto c_i c_j Q_{ij}\) and hence \(V_{ij} \mapsto V_{ij}/(c_i c_j)\), which cancels exactly in the ratio \(R_{ij} = V_{ij}/\sqrt{V_{ii}V_{jj}}\), leaving \(R(\theta)\) unchanged.
Here, the \(d_i\)'s act as scale factors on the free parameters \(\theta\), affecting the numerical conditioning of the forward map and the efficiency with which samplers or optimisers traverse the parameter space.
The choice \(d_i = p - i + 1\) is a natural decreasing sequence that keeps the Cholesky entries of comparable scale.
This particular sequence is not arbitrary: it coincides with the expected squared Cholesky diagonals of a minimally proper Wishart matrix, a connection to the Bartlett decomposition that we develop in the Appendix.
\end{remark}

With the GCP as the bijection (\ref{eq-rtheta}), the priors (\ref{eq-prior-xi}) and (\ref{eq-prior-theta}) apply at dimension \(m = n_Q\).
The remaining ingredient is the Hessian \(H(\theta_0)\) of (\ref{eq-hessian}), which requires the base-model parameter \(\theta_0\) corresponding to a user-supplied \(R_0\).
Its computation is the subject of Section~\ref{sec-base-conversion} below.

\subsection{Properties of the GCP Map}\label{properties-of-the-gcp-map}

Let
\begin{equation}\protect\phantomsection\label{eq-rpg}{
\mathcal{R}_p(\mathcal{G}) \;:=\; \big\{ R \in \mathcal{R}_p : R^{-1} \text{ is } \mathcal{G}\text{-sparse} \big\}
}\end{equation}
denote the submanifold of correlation matrices whose precision respects the conditional independence structure of \(\mathcal{G}\).
We record the following properties for convenience; most follow immediately from the construction or from standard facts about Cholesky factorisation.

\begin{proposition}[Properties of the GCP map]\protect\hypertarget{prp-gcp}{}\label{prp-gcp}

For every \(\theta \in \mathbb{R}^{n_Q}\), the GCP map \(\theta \mapsto R(\theta)\) satisfies:

\begin{enumerate}
\def\labelenumi{(\roman{enumi})}
\item
  \emph{\textbf{Validity and \(\mathcal{G}\)-sparsity.}} \(R(\theta) \in \mathcal{R}_p(\mathcal{G})\), i.e.~\(R(\theta) \succ 0\) and \(R(\theta)^{-1}\) is \(\mathcal{G}\)-sparse.
\item
  \emph{\textbf{Smoothness.}} \(\theta \mapsto R(\theta)\) is real-analytic on \(\mathbb{R}^{n_Q}\).
\item
  \emph{\textbf{Bijectivity.}} \(\theta \mapsto R(\theta)\) is a bijection from \(\mathbb{R}^{n_Q}\) onto \(\mathcal{R}_p(\mathcal{G})\); the inverse \(R \mapsto \theta\) is defined only on \(\mathcal{R}_p(\mathcal{G})\), which is a strict submanifold of \(\mathcal{R}_p\) whenever \(\mathcal{E} \subsetneq \binom{\mathcal{V}}{2}\).
\end{enumerate}

\end{proposition}

To show part (i), reason the following: the strictly positive diagonal of \(L(\theta)\) gives \(Q(\theta) = L(\theta) L(\theta)^{\top} \succ 0\), while (\ref{eq-gcp-fillin}) and Lemma~\ref{lem-prelim-three-class}(i) zero every non-edge entry of \(Q(\theta)\), so the normalisation (\ref{eq-forward-map}) returns \(R(\theta) \in \mathcal{R}_p(\mathcal{G})\).
Validity and \(\mathcal{G}\)-sparsity thus hold automatically, for every \(\theta \in \mathbb{R}^{n_Q}\) and with no constraint on the parameters.
By contrast, Cholesky-based parameterisations that factor \(R = LL^\top\), notably the LKJ construction of \citet{lewandowski2009generating}, must instead confine the off-diagonal entries of \(L\) to a complicated, \(p\)-dependent feasible set to keep \(R\) positive definite.

Parts (ii) and (iii) concern the map's analytic structure.
For (ii), each entry of \(L(\theta)\) is polynomial in \(\theta\) up to the constant factors \(d_j\), and matrix inversion and diagonal normalisation are smooth on \(\mathcal{S}_{++}^{p}\), so \(\theta \mapsto R(\theta)\) is real-analytic.
Bijectivity (iii) follows from the uniqueness of the Cholesky factorisation, and is in fact the correlation-scale counterpart of the bijection between \(\mathcal{S}_{++}^{p}(\mathcal{G})\) and its free Cholesky entries established by \citet[App. A]{roverato2002hyper} and \citet[Sec. 2]{atay2005monte}.
The precision \(R^{-1}\) of any \(R \in \mathcal{R}_p(\mathcal{G})\) is \(\mathcal{G}\)-sparse with a unique Cholesky factor, whose edge entries, once rescaled to the fixed diagonal of Section~\ref{sec-base-conversion}, recover a unique \(\theta\) via Lemma~\ref{lem-prelim-three-class}.
The inclusion \(\mathcal{R}_p(\mathcal{G}) \subsetneq \mathcal{R}_p\) is strict whenever \(\mathcal{E} \subsetneq \binom{\mathcal{V}}{2}\), as a generic correlation matrix has dense inverse.
Both properties are needed to build the prior.
Smoothness supplies the Hessian \(H(\theta_0)\) that whitens the geometry, while the bijection lets the prior on \(\theta\) be transported unambiguously to a prior on \(R\) supported on all of \(\mathcal{R}_p(\mathcal{G})\).

\begin{remark}[Vertex ordering]
Each fill-in entry is a rational function of \(\theta\) via (\ref{eq-gcp-fillin}) and therefore contributes non-linear terms to the Jacobian of \(\theta \mapsto R(\theta)\) needed when evaluating \(H(\theta_0)\).
Since the recursion is columnwise, the nonlinearity compounds along chains of fill-ins.
As discussed in Section~\ref{sec-fillin}, the fill-in count \(|\mathcal{I}_{\mathrm{fill}}| = |\mathcal{E}^*| - n_Q\) depends on the vertex labelling and can be reduced (sometimes to zero for chordal graphs) by applying a suitable reordering of \(\mathcal{V}\) before constructing the GCP.
Fewer fill-in entries therefore mean a milder nonlinearity in \(\theta \mapsto R(\theta)\), so the second-order approximation on which the prior rests is correspondingly more accurate.
\end{remark}

\subsection{Reference Model}\label{sec-base-conversion}

The prior requires the base-model parameter \(\theta_0\) corresponding to a user-specified correlation matrix \(R_0 \in \mathcal{R}_p\).
The inverse map \(R_0 \mapsto \theta_0\) proceeds as follows.

% \textbf{Algorithm} (Base Model Conversion). Given \(R_0 \in \mathcal{R}_p\) and \(\mathcal{G}\):

% \begin{enumerate}
% \def\labelenumi{\arabic{enumi}.}
% \tightlist
% \item
%   Compute \(Q_0 = R_0^{-1}\) and verify that \((Q_0)_{ij} = 0\) for all \((i,j) \notin \mathcal{E}\).

%   \begin{itemize}
%   \tightlist
%   \item
%     If not, \(R_0\) must be adjusted or \(\mathcal{G}\) extended.
%   \end{itemize}
% \item
%   Compute the Cholesky factorisation \(Q_0 = \hat{L}\hat{L}^\top\).
% \item
%   Rescale each row \(i\) to enforce fixed diagonals \(d_i\), i.e.~set \(\tilde{L}_{ij} = \hat{L}_{ij}\, d_i\, / \hat{L}_{ii}\) for \(j \le i\).
% \item
%   Extract \(\theta_0 = \{\tilde{L}_{ij} : i > j,\, \{i,j\} \in \mathcal{E}\}\).
% \end{enumerate}

\begin{algorithm}
\caption{(Reference Correlation). Given \(R_0 \in \mathcal{R}_p\) and \(\mathcal{G}\)}
\label{alg:algorithm1}
\begin{algorithmic}[1]
\State  Compute \(Q_0 = R_0^{-1}\) and verify that \((Q_0)_{ij} = 0\) for all \((i,j) \notin \mathcal{E}\).    If not, \(R_0\) must be adjusted or \(\mathcal{G}\) extended.
\State  Compute the Cholesky factorisation \(Q_0 = \hat{L}\hat{L}^\top\).
\State Rescale each row \(i\) to enforce fixed diagonals \(d_i\), i.e.~set \(\tilde{L}_{ij} = \hat{L}_{ij}\, d_i\, / \hat{L}_{ii}\) for \(j \le i\).
\State  Extract \(\theta_0 = \{\tilde{L}_{ij} : i > j,\, \{i,j\} \in \mathcal{E}\}\).
\end{algorithmic}
\end{algorithm}

The common choice \(R_0 = I_p\) gives \(\theta_0 = 0\) under any graph, since \(I_p^{-1} = I_p\) and the Cholesky factor is diagonal.
More generally, the rescaling in Step 3 guarantees \(R(\theta_0) = R_0\): multiplying row \(i\) of \(\hat{L}\) by \(d_i/\hat{L}_{ii}\) amounts to a diagonal congruence transformation of \(Q_0\), which is exactly cancelled by the normalisation in (\ref{eq-forward-map}), leaving \(R(\theta_0)\) unchanged.

This procedure also makes the asymmetry of Proposition~\ref{prp-gcp}(iii) operationally transparent.
The forward map accepts any \(\theta \in \mathbb{R}^{n_Q}\) and always returns a \(\mathcal{G}\)-compatible correlation matrix, so optimisers and MCMC samplers may explore the parameter space freely.
On the other hand, Step 1 reveals that the inverse is defined only on \(\mathcal{R}_p(\mathcal{G}) \subsetneq \mathcal{R}_p\), so an arbitrary correlation matrix such as an empirical estimate cannot be mapped back to \(\theta\) \emph{unless} its precision already carries the sparsity pattern of \(\mathcal{G}\).
A reference model that fails this must be reconciled with \(\mathcal{G}\) before the prior can be centred on it.
This entails extending \(\mathcal{G}\) to cover the non-zero off-diagonals of \(Q_0\), at worst to the complete graph (the dense case), which accommodates every \(R_0\).

The reference parameter \(\theta_0\) and the Hessian \(H(\theta_0)\) are computed once, offline.
They are then reused wherever the prior is evaluated, such as drawing from it with the proposed sampler described in the next section, or to compute its log-density within posterior maximisation and MCMC schemes.

\subsection{Sampling from the Prior}\label{sampling-from-the-prior}

Although the prior on \(\theta\) is available in closed form (\ref{eq-prior-theta}), the induced density on \(R\) is awkward to evaluate and, for \(m \ge 2\), unbounded at the mode (Proposition~\ref{prp-properties}), so drawing samples is the practical route to inspecting or plotting it.
Exact draws are obtained by inverting the whitening transformation (\ref{eq-whitening}).
The marginal distribution of \(r = \|\xi\|\) is \(\operatorname{Exp}(\lambda)\) and, conditional on \(r\), the direction \({u} = \xi/r\) is uniform on \(\mathbb{S}^{m-1}\) by (\ref{eq-prior-spherical}).
The two components are independent and each straightforward to simulate.

\begin{algorithm}
\caption{Prior draw for $R$. Given \(\theta_0\), \(H(\theta_0)\), and \(\lambda > 0\)}
\label{alg:algorithm2}
\begin{algorithmic}[1]
\State Draw $r \sim \operatorname{Exp}(\lambda)$.
\State Draw $z \sim N_m(0,I_m)$ independently, and set
       $u=z/\|z\|$ and $\xi=ru$.
\State Affine transform
       $\theta = H(\theta_0)^{-1/2}\xi+\theta_0$.
\State Compute $R$ via the GCP map $\theta \mapsto R(\theta)$.
\end{algorithmic}
\end{algorithm}

Step 1 exploits the fact that \(r \sim \operatorname{Exp}(\lambda)\) can be drawn as \(r = -\log(U)/\lambda\) for \(U \sim \operatorname{Uniform}(0,1)\).
Step 2 uses the standard result that \(z/\|z\|\) is uniform on \(\mathbb{S}^{m-1}\) whenever \(z\) is spherically symmetric, so any \(\operatorname{N}_m({0}, I_m)\) draw normalised to unit length suffices.
Both steps are straightforward because, in the whitened coordinates \(\xi\), the prior density is isotropic.
As a note, when \(\mathcal{G}\) is the complete graph, the GCP map reduces to the dense parameterisation of Section~\ref{sec-prior}, and the same four steps apply without modification.
Prior predictive draws, sensitivity analyses with respect to \(\lambda\), and direct verification that the elicited prior is consistent with domain knowledge are then possible.

\subsection{Illustration}\label{sec-illustration-sparse}

We illustrate the parameterisation for \(p = 3\) under the graph \(\mathcal{G}\) with edges \((1,2)\) and \((1,3)\) but not \((2,3)\), encoding the conditional independence \(x_2 \perp x_3 \mid x_1\).
The precision matrix \(Q\) has the lower-triangular sparsity pattern induced by \(\mathcal{G}\), whose filled graph \(\mathcal{G}^{*}\) is shown below.

\begin{center}
\includegraphics[width=0.5\linewidth,height=\textheight,keepaspectratio]{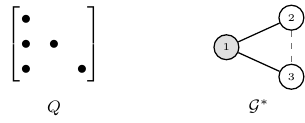}
\end{center}

The Cholesky factor, with fixed diagonal \((d_1, d_2, d_3) = (3, 2, 1)\), is formed as
\begin{equation}\protect\phantomsection\label{eq-L-illustration}{
L(\theta) = \begin{pmatrix} 3 & & \\ \theta_1 & 2 & \\ \theta_2 & \ell_{32} & 1 \end{pmatrix}, \qquad
\ell_{32} = -\frac{\theta_1 \theta_2}{2},
}\end{equation}
where \(\ell_{32}\) is the unique fill-in entry, determined analytically from \((\theta_1, \theta_2)\) by (\ref{eq-gcp-fillin}).
The correlation matrix \(R(\theta_1, \theta_2)\) obtained from (\ref{eq-forward-map}) has three off-diagonal entries, all expressed as functions of only two parameters.

Consider first the base model \(R_0 = I_3\) (independence), giving \(\theta_0 = (0,0)^\top\).
The Hessian \(H(\theta_0)\) at the origin reduces to a \(2 \times 2\) matrix determined by the graph structure, here admitting the closed form \(\operatorname{diag}(1/4,\, 1)\), while in general it is computed once offline by numerical differentiation.
The prior (\ref{eq-prior-theta}) then concentrates mass exponentially in the Euclidean norm \(r = \|H(\theta_0)^{1/2}\theta\|\).
Each unit of \(r\) is one Fisher distance unit away from \(R_0\), and the rate \(\lambda\) controls the prior expected distance through \(\operatorname{E}[r] = 1/\lambda\).

The base model need not be independence; any \(R_0\) that is itself Markov to \(\mathcal{G}\) may anchor the prior.
Because \(\mathcal{G}\) is a tree, this admissibility forces the leaf correlation to factorise along the path, \(\rho_{23} = \rho_{12}\rho_{13}\), so the base is pinned down by its two hub correlations alone.
Fixing \(\rho_{12} = -0.5\) and \(\rho_{13} = 0.7\) therefore sets \(\rho_{23} = -0.35\), and the Base Model Conversion algorithm of Section~\ref{sec-base-conversion} returns the corresponding base parameter \(\theta_0 = (0.825,\, -0.906)^\top\).
The construction is otherwise unchanged with respect to the independence base.
What's changed is that \(H(\theta_0)\) is no longer diagonal, though the prior remains isotropic in the whitened coordinate and shrinks toward \(R_0\) at rate \(\lambda\).

\protect\phantomsection\label{cell-fig-gcp-prior-illustration}
\begin{figure}[p]

\centering{

\includegraphics[width=1\linewidth,height=\textheight,keepaspectratio]{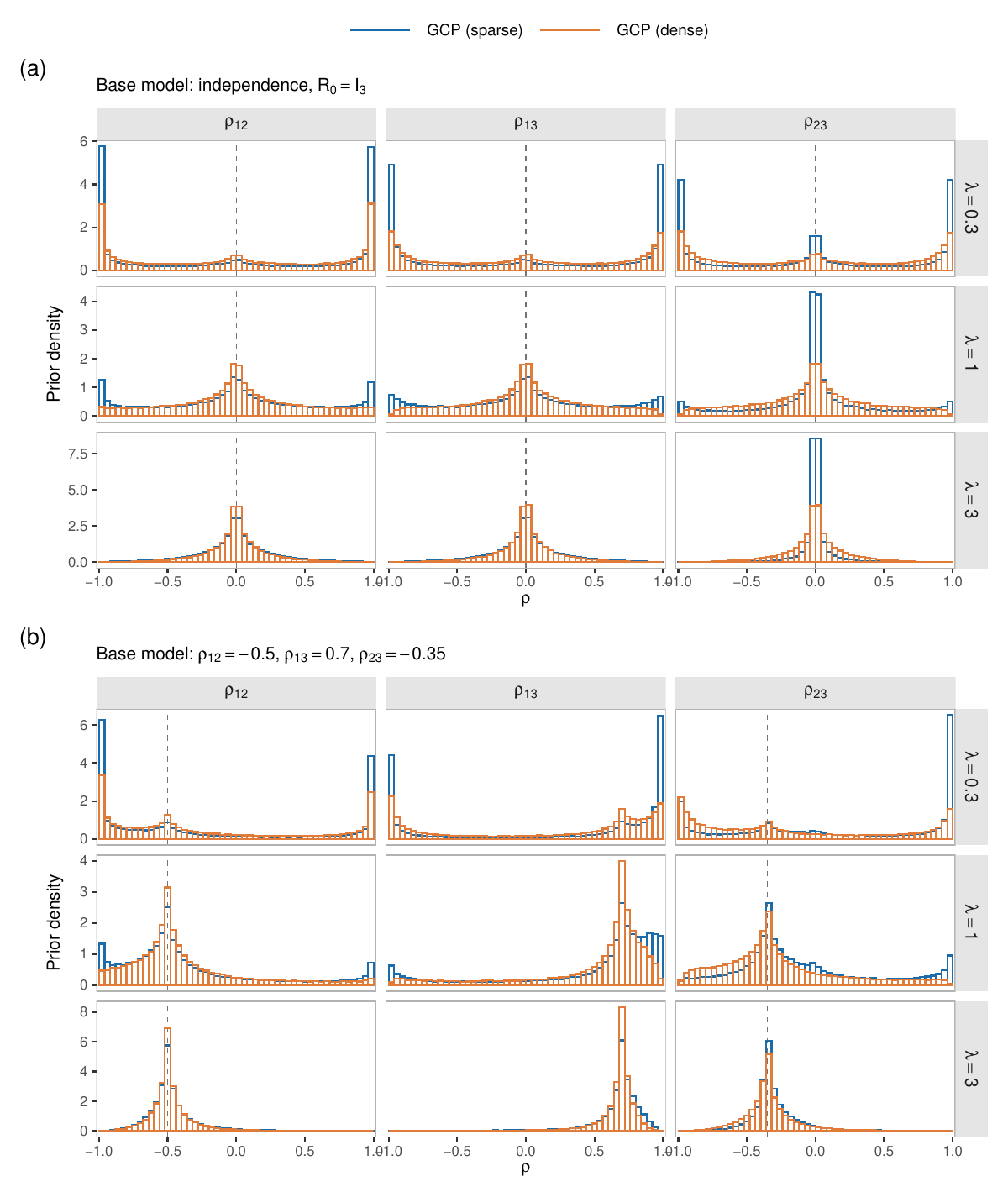}

}

\caption{\label{fig-gcp-prior-illustration}Marginal prior densities for the three pairwise correlations \((\rho_{12}, \rho_{13}, \rho_{23})\) under the GCP prior with \(\mathcal{G}\) a star graph (blue) and with \(\mathcal{G}\) the complete graph (orange), shown for \(\lambda \in \{0.3, 1, 3\}\). Panel (a) takes the base model to be independence, \(R_0 = I_3\); panel (b) takes a non-identity base in which the correlations are set to \(\rho_{12} = -0.5\), \(\rho_{13} = 0.7\), and \(\rho_{23} = \rho_{12}\rho_{13} = -0.35\). Dashed lines mark the base correlations. All densities are histograms based on \(N = 50{,}000\) prior samples.}

\end{figure}%

Figure~\ref{fig-gcp-prior-illustration} shows the resulting marginals for the two base models under both this sparse graph (blue) and, for comparison, the complete graph (orange).
Although \(\lambda\) controls concentration uniformly about the origin in the whitened parameter space, its expression in correlation space is shaped by the forward map.
In panel (a) with the independence base, the free correlations \(\rho_{12}\) and \(\rho_{13}\)---each a single \(\theta_j\) pushed through a saturating map onto \((-1, 1)\)---contract symmetrically toward zero as \(\lambda\) grows and spread toward \(\pm 1\) as it shrinks.
In panel (b) with the non-identity base, they concentrate instead on the base values marked by the dashed lines, \((-0.5, 0.7)\).
Most notably, the free pair reproduces the boundary behaviour established for the univariate prior in Section~\ref{sec-prior}: the small-\(\lambda\) migration of mass toward \(\pm 1\), seen as the near-boundary spikes in the outer histogram bins.
This mirrors the univariate spread of Figure~\ref{fig-uni-prior}.

The genuinely multivariate feature is the coupled correlation \(\rho_{23}\), which is not directly parameterised but enters through the fill-in \(\ell_{32}\) as a function of the product \(\theta_1\theta_2\).
It is therefore tighter than the free pair at every \(\lambda\) and tracks \(\rho_{12}\rho_{13}\), settling at \(-0.35\) in panel (b).
The complete graph instead leaves \(\ell_{32}\) free, severing this tie so that \(\rho_{23}\) disperses as widely as the others, the blue-orange gap being exactly the prior mass that the conditional independence \(x_2 \perp x_3 \mid x_1\) removes.
Crucially, contraction holds for the \emph{derived correlation} no less than for the free pair: as \(\lambda\) grows, all three marginals concentrate on their base values.
The prior thus contracts onto the user-specified \(R_0\) as a whole, and the proposed prior's defining contraction property is now realised across a full correlation matrix rather than a single parameter.

\section{Illustrative Example: Heart Disease in South Africa}\label{sec-data}

We analyse a retrospective sample of males in a heart-disease high-risk region of the Western Cape, South Africa \citep{rossouw1983coronary}. The data contain \(n = 462\) observations on the 10 variables listed in Table~\ref{tbl-saheart}, i.e.~eight clinical and behavioural measurements, a binary indicator of family history, and the binary response for coronary heart disease.

\begin{table}[htbp]

\caption{\label{tbl-saheart}Description of the 10 variables in the South African heart disease data.}

\centering{

\centering

\vspace{1em}
\begin{tabular}{lll}
\hline
 & Variable & Description \\
\hline
$x_1$ & \texttt{sbp}       & Systolic blood pressure \\
$x_2$ & \texttt{ldl}       & Low density lipoprotein cholesterol \\
$x_3$ & \texttt{adiposity} & Accumulation of excess fatty tissue \\
$x_4$ & \texttt{typea}     & Type-A behaviour \\
$x_5$ & \texttt{obesity}   & Obesity index \\
$x_6$ & \texttt{age}       & Age at onset of study \\
$x_7$ & \texttt{tobacco}   & Cumulative tobacco use (kg) \\
$x_8$ & \texttt{alcohol}   & Current alcohol consumption \\
$x_9$ & \texttt{famhist}   & Family history of heart disease (binary) \\
$y$   & \texttt{chd}       & Coronary heart disease (response, binary) \\
\hline
\end{tabular}

}
\end{table}%

Our aim is to characterise the dependence among the nine covariates \(x_1, \dots, x_9\) and to carry that structure into a model for the binary response \(y\) (\texttt{chd}).
We model each covariate as a link-transformed manifestation of a 9-dimensional standardised latent Gaussian field \(x^*\), whose \(9 \times 9\) correlation matrix \(R\) carries all the dependence and receives our proposed prior.
For each subject \(s = 1, \dots, n\), the generative model is
\begin{equation}\protect\phantomsection\label{eq-saheart-model}{
\begin{gathered}
y_s \sim \mathrm{Bernoulli}\big(\operatorname{logit}^{-1}(\beta_0 + x_s^{*\top}\beta)\big), \\ \qquad
x_{si} \begin{cases}
  = \mu_i + \sigma_i\, x^*_{si} & i = 1, \dots, 6, \\
  \sim \mathrm{Gamma}\big(\phi_i,\, \phi_i\, e^{-(a_i + x^*_{si})}\big) & i = 7, 8, \\
  \sim \mathrm{Bernoulli}\big(\Phi(\alpha + x^*_{si})\big) & i = 9,
\end{cases} \\
x^*_s \sim \mathrm{N}_9(0, R),
\end{gathered}
}\end{equation}
where \(\Phi\) is the standard normal distribution function, the Gamma is written in shape--rate form so that covariate \(i\) has mean \(\exp(a_i + x^*_{si})\) and dispersion \(\phi_i\), and the location, scale, and regression hyperparameters all carry weakly informative priors.
The six approximately symmetric continuous measurements are linked to their latent coordinates by a location-scale transform with freely estimated mean \(\mu_i\) and scale \(\sigma_i\), while the two skewed positive measurements enter through a log-mean Gamma model with its own dispersion, and the binary \texttt{famhist} through a probit link.
The response \texttt{chd} is a logistic regression on the same latent coordinates, so that the covariate dependence and the outcome model are estimated jointly rather than in sequence.
All dependence among the covariates thus resides in \(R\), the correlation matrix of the latent field \(x^*\), and it is on \(R\) that the proposed prior acts.
Inference is carried out in Stan \citep{carpenter2017stan} with the prior supplied through the reparameterisation \(\theta \mapsto R(\theta)\) so that sampling proceeds over the unconstrained edge parameters \(\theta \in \mathbb{R}^{n_Q}\).

\begin{figure}[!htbp]

\centering{

\includegraphics[width=4.9875in,height=\textheight,keepaspectratio]{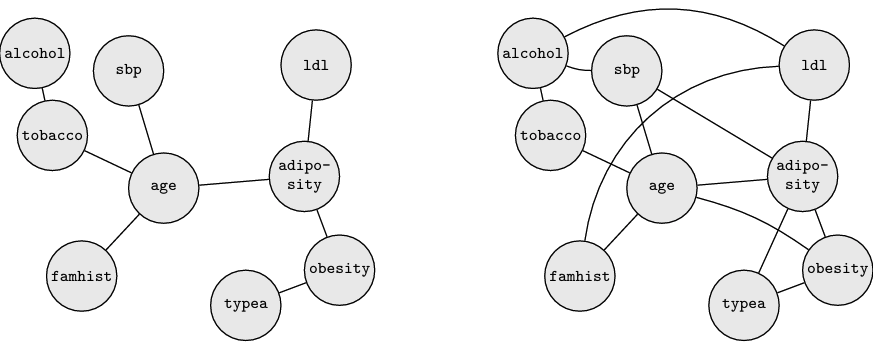}

}

\caption{\label{fig-saheart-tikz}Two dependency graphs on the nine covariates of the South African heart data, $\mathcal{G}_{1}$ (left) with 8 edges, and the denser graph $\mathcal{G}_{2}$ (right).}

\end{figure}

The proposed prior is indexed by a graph \(\mathcal{G}\) on the nine covariates, which, in practice, the user supplies together with a correlation \(R_0\) that is Markov to \(\mathcal{G}\).
Here we consider two possible graphs and examine how this choice shapes the resulting posterior. These graphs are constructed from expert opinions on the risk factors for coronary heart disease and the interfactor dependence (see \citet{kong2024bayesian, li2017sleep, micha2010red} for details).
%Writing \(q_{ij}\) for the empirical partial correlation between covariates \(i\) and \(j\), we form (i) a minimum spanning tree (MST) on the weights \(w_{ij} = 1 - |q_{ij}|\), which links all nine nodes at the least total weight and so retains the strongest partial correlations, giving a tree with eight edges; and (ii) a thresholded graph that instead places an edge wherever \(|q_{ij}| > 0.07\), giving 14 edges.
Both are shown in Figure~\ref{fig-saheart-tikz}. \\

Now we have to construct an $R_0$. Based on the graphs, we can assign equal partial correlations to all edges in the absence of additional information. Alternatively, from the literature on coronary heart disease, $R_0$ can be defined based on the posterior correlation matrices of previous studies. For illustration, we use the first approach, where all partial correlations for the edges in the edge set is considered as $0.2$. We consider a common rate parameter \(\lambda = 1\).

\protect\phantomsection\label{cell-fig-saheart-tikz}

\begin{figure}[hp]

\centering{

\includegraphics[width=1\linewidth,height=\textheight,keepaspectratio]{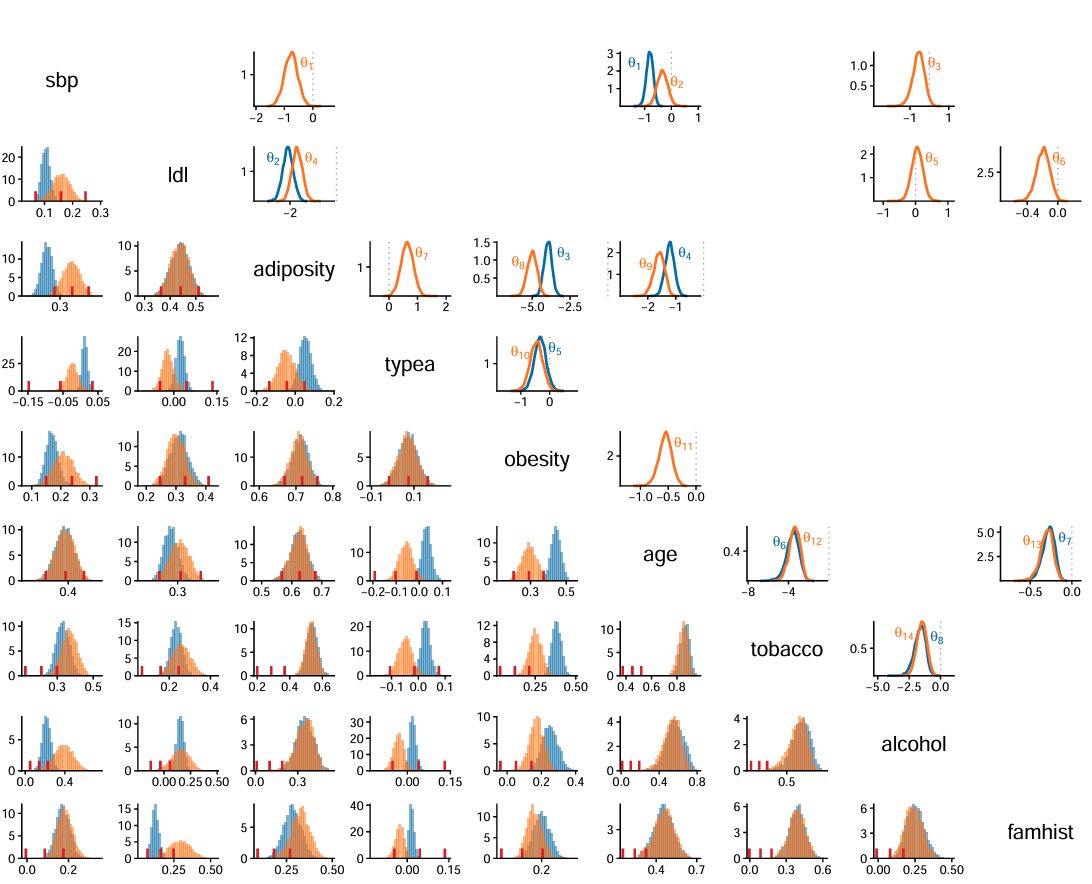}

}

\caption{\label{fig-saheart-results}Posterior summaries for the South African heart data under the proposed prior with graph $\mathcal{G}_{1}$ (blue) and graph $\mathcal{G}_{2}$ (orange). Lower-triangle panels give the posterior marginal of each pairwise correlation as overlaid histograms, with the observed correlation and its Fisher confidence interval marked in red. Upper-triangle panels give the posteriors of the GCP parameters \(\theta\) as smooth kernel density estimates of the MCMC samples, each curve labelled with its \(\theta\) index.}

\end{figure}

Each graph reduces the \(\binom{9}{2} = 36\) free correlations to its edge count: 8 for $\mathcal{G}_{1}$ and 14 for $\mathcal{G}_{2}$.
Figure~\ref{fig-saheart-results} shows the posteriors they induce, with each observed correlation and its Fisher confidence interval marked for reference.
The prior's structure can be read off directly.
Because $\mathcal{G}_{1}$ is a tree, any pair of covariates not joined by an edge carries no parameter of its own, so the prior factorises its correlation along the unique connecting path---the leaf-correlation behaviour of Section~\ref{sec-illustration-sparse}, now across a \(9 \times 9\) matrix rather than a single triple.
In $\mathcal{G}_{2}$ this is relaxed, whereby each added edge severs one conditional independence and frees the correlation it had constrained, while the pairs left non-adjacent stay factorised.
This contrast is visible in Figure~\ref{fig-saheart-results}, though across most of the matrix the two posteriors are close.
They coincide on the eight edges common to both graphs and on many of the pairs left non-adjacent by both, separating mainly on the six edges that is added to the tree through $\mathcal{G}_{2}$.
Here, the extra parameters free the correlation to follow the data, whereas $\mathcal{G}_{1}$, lacking the extra edges, reports the value implied by the connecting path, which may understate the direct association, as at \texttt{sbp}-\texttt{adiposity}, or overstate it, as at \texttt{obesity}-\texttt{age}.
For the Gaussian covariates, the freed correlation settles at the observed value.
Pairs that involve a gamma or probit variable tend instead to sit above their observed marginals, the nonlinear link attenuating the association the raw data record, so the red markers serve there only as a reference.

\protect\phantomsection\label{cell-fig-saheart-results}

% Flush the [p] float page at the next page boundary (no hard mid-page break).
% Source - https://tex.stackexchange.com/a/247935
% Posted by hadi
% Retrieved 2026-06-07, License - CC BY-SA 3.0
\afterpage{\FloatBarrier}

\section{Conclusion}\label{sec-conclusion}

We have proposed an informative, distance-based prior for correlation matrices that shrinks toward a user-specified reference, resolving two issues persisting from previous work:
(a) the inability to centre the prior on a substantively motivated correlation structure; and (b) the restriction to a fully unstructured parameterisation when conditional independence constraints are available.
The construction rests on three user-specified ingredients---a reference correlation matrix \(R_0\), a graph \(\mathcal{G}\) encoding conditional independence, and a scalar rate \(\lambda\)---and assigns mass that decays exponentially in the Fisher arc-length distance from \(R_0\).
Enabling this is the Graphical Cholesky Parameterisation, which reduces the parameter dimension from \(\binom{p}{2}\) to \(|\mathcal{E}|\) while preserving the bijectivity and smoothness the whitening construction requires.
The resulting prior accommodates both positive and negative correlations, is proper for all \(\lambda > 0\), and recovers the fully unstructured dense case when \(\mathcal{G}\) is the complete graph.

In practice, the method is suited to moderate dimensions, roughly \(p \leq 150\), where the GCP map and the Fisher information Hessian are computationally tractable.
Helper functions implementing both the dense and sparse priors are available in the \texttt{graphpcor} package \citep{krainski2026graphpcor}, which supports prior specification and sampling in Stan \citep{carpenter2017stan} and R-INLA \citep{rue2009approximate}.

We encourage users to conduct prior predictive checks \citep{gabry2019visualization} before committing to a particular choice of \(R_0\), \(\mathcal{G}\), and \(\lambda\).
The rate parameter \(\lambda\) lacks a direct subject-matter interpretation, so calibrating it to a meaningful prior belief about departure from \(R_0\) is not immediate.
Prior predictive simulation provides a principled route to this calibration.
The sampling algorithm developed in Section~\ref{sec-sparse} generates prior draws on \(\mathcal{R}_p\), and the implied distributions over individual correlations or derived quantities can be inspected and compared against domain knowledge before any data are observed.

One practical consideration deserves emphasis.
The graph \(\mathcal{G}\) must be compatible with the base model, i.e.~the precision \(Q_0 = R_0^{-1}\) must carry the sparsity pattern of \(\mathcal{G}\).
If it does not, \(\mathcal{G}\) must be extended to accommodate the non-zero entries of \(Q_0\), reverting to the complete graph in the worst case.
Selecting \(\mathcal{G}\) to reflect genuine conditional independence structure is a substantive modelling decision.
Many algorithms exist for this purpose, though graph selection itself falls outside the scope of this paper.

\section*{Supplementary Material}\label{supplementary-material}
\addcontentsline{toc}{section}{Supplementary Material}

All code and data required to reproduce the analyses and figures in this paper are provided on the GitHub  at \url{https://github.com/alan-turing-institute/InformativePrior}.
{An appendix containing the proofs of Lemma~\ref{lem-prelim-three-class} and Proposition~\ref{prp-properties} together with a remark regarding the connection to the Bartlett decomposition, is provided as separate supplementary material at the same URL.}

\section*{Appendix}\label{appendix}
\addcontentsline{toc}{section}{Appendix}

\subsection*{\texorpdfstring{Proof of Three-Class Structure of
\(L\)}{Proof of Three-Class Structure of L}}\label{proof-of-three-class-structure-of-l}
\addcontentsline{toc}{subsection}{Proof of Three-Class Structure of
($L$)}

All three parts are established by induction on the column index \(j\).
The base case \(j = 1\) is immediate: \(L_{i1} = Q_{i1}/L_{11}\), which
is non-zero only when \((i,1) \in \mathcal{I}_{\mathrm{edge}}\), and
depends only on \(Q_{i1}\) and \(Q_{11}\).

For the inductive step at column \(j \geq 2\), consider any \(i > j\).
Since \(Q_{ij} = 0\) for \((i,j) \notin \mathcal{I}_{\mathrm{edge}}\),
{equation (\ref{eq-prelim-chol})} reduces to \[
L_{ij} = -L_{jj}^{-1} \sum_{k < j} L_{ik}\,L_{jk}.
\] By induction, each factor \(L_{ik}\) and \(L_{jk}\) with \(k < j\) is
already determined by
\(\{Q_{rs} : (r,s) \in \mathcal{I}_{\mathrm{edge}},\, s < k\}\) and
\(\{Q_{ll} : l \leq k\}\), establishing (ii) and (iii).

For (i), suppose \((i,j) \in \mathcal{I}_{\mathrm{zero}}\) and the sum
is non-zero; then there exists \(k < j\) with \(L_{ik} \neq 0\) and
\(L_{jk} \neq 0\). By the inductive hypothesis,
\(\{i,k\} \in \mathcal{E}^{*}\) and \(\{j,k\} \in \mathcal{E}^{*}\). By
{(\ref{def-prelim-filled})}, each is certified by a path in
\(\mathcal{G}\) whose interior vertices lie strictly below \(k\).
Concatenating these paths gives a path from \(i\) to \(j\) with all
interior vertices at most \(k < j = \min(i,j)\), so
\(\{i,j\} \in \mathcal{E}^{*}\), contradicting
\((i,j) \in \mathcal{I}_{\mathrm{zero}}\). Hence \(L_{ij} = 0\).
\(\square\)

\subsection*{Proof of Contraction
Property}\label{proof-of-contraction-property}
\addcontentsline{toc}{subsection}{Proof of Contraction Property}

Write
\(\pi(\xi ; \lambda) = c_m \, \lambda\, e^{-\lambda r}\, r^{-(m-1)}\)
with \(r = \|\xi\|\) and \(c_m = \Gamma(m/2)/(2\pi^{m/2})\).

\emph{Strict monotonicity in \(r\).} Taking logarithms,
\(\log\pi(\xi ; \lambda) = \log(c_m \lambda) - \lambda r - (m-1)\log r\),
whose derivative with respect to \(r\) is \(-\lambda - (m-1)/r < 0\) for
all \(r > 0\) and \(m \ge 1\). Hence the density is strictly decreasing
in \(r\), and level sets \(\{\xi : \pi(\xi ; \lambda) \ge c\}\) are
closed balls centred at the origin (equivalently, centred at
\(\theta_0\) under {Equation~\ref{eq-whitening}}).

\emph{Behaviour at the origin.} As \(r \to 0^+\),
\(e^{-\lambda r} \to 1\) and \(r^{-(m-1)} \to 1\) if \(m = 1\) or
\(+\infty\) if \(m \ge 2\). Thus for \(m = 1\) the density attains the
finite maximum \(\pi(0 ; \lambda) = \lambda/2\) (the Laplace
distribution), whereas for \(m \ge 2\) the density diverges at
\(\xi = 0\). In the latter case the singularity is integrable: near the
origin the radial contribution is
\(\int_0^\varepsilon r^{-(m-1)} \cdot r^{m-1}\, dr = \varepsilon < \infty\),
consistent with the propriety established in part (i).

\emph{Geometric interpretation.} Under {Equation~\ref{eq-whitening}} the
level sets of the prior on \(\theta\) are ellipsoids
\(\{\theta : (\theta - \theta_0)^\top H(\theta_0) (\theta - \theta_0) \le c^2\}\),
which, by {Equation~\ref{eq-kld-taylor}}, approximate the KL-divergence
sublevel sets
\(\{\theta : \operatorname{D}_{\mathrm{KL}}(R(\theta) \Vert R_0) \le c^2/2\}\)
near \(\theta_0\). The prior therefore favours \(R\) that are close to
the base model in KLD, with concentration controlled by \(\lambda\).
This completes the proof.

\subsection*{Comparison to the LKJ and Wishart
Priors}\label{comparison-to-the-lkj-and-wishart-priors}
\addcontentsline{toc}{subsection}{Comparison to the LKJ and Wishart
Priors}

Fix the base model at the identity, \(R_0 = I_p\). The Kullback-Leibler
divergence then reduces to
\(\operatorname{D}_{\mathrm{KL}}(\operatorname{N}(0, R) \,\Vert\, \operatorname{N}(0, I_p)) = -\tfrac{1}{2}\log|R|\),
so the squared Fisher arc length is
\(d^2 = 2\operatorname{D}_{\mathrm{KL}} = -\log|R|\). The LKJ prior of
\textcite{lewandowski2009generating} may therefore be written in the
same geometry, \[
\pi_{\mathrm{LKJ}}(R) \propto |R|^{\eta-1} = \exp\!\big(-(\eta-1)\,d^2\big),
\] a Gaussian decay in the Fisher arc length from the identity. The
prior of the main article shares this base model but, by the
constant-rate principle {(\ref{eq-pc-rate})}, replaces the quadratic
penalty \((\eta - 1)d^2\) with the linear penalty \(\lambda d\), giving
the exponential decay \(\exp(-\lambda d)\). Both priors contract toward
the same \(I_p\); they differ only in how departures from it are
penalised.

The two penalties also behave differently at the boundary of
\(\mathcal{R}_p\). Take the two-dimensional case \(p = 2\). Pushed
forward to a single correlation \(\rho \in (-1, 1)\), the prior of the
main article is an exponentially decaying kernel in the unconstrained
coordinate multiplied by a polynomially growing Jacobian. As
\(\rho \to \pm 1\) the exponential dominates and the density falls to
zero for every \(\lambda > 0\). The LKJ, beta, and Wishart-implied
marginals instead decay polynomially,
\(\pi(\rho) \propto (1 - \rho^2)^{c}\) (with \(c = \eta - 1\) for LKJ),
and so diverge at the boundary whenever \(c < 0\).

Finally, the constructions treat dimension differently. The radial
density underlying the prior of the main article is normalised by the
surface area \(2\pi^{m/2}/\Gamma(m/2)\) of the unit sphere
\(\mathbb{S}^{m-1}\), so that \(\lambda\) retains a fixed interpretation
as the exponential rate of the Fisher distance \(r = \|\xi\|\)
irrespective of the dimension \(m\). The LKJ shape parameter \(\eta\)
admits no such normalisation. If \(\eta\) is kept constant, the
distribution concentrates more tightly around \(I_p\) as the dimension
\(p\) increases \autocite{tokuda2025visualizing}. Because of this,
\(\eta\) cannot represent a consistent belief about the distance from
the target across different dimensions. The Wishart degrees of freedom
suffer from this exact same limitation.

\subsection*{Connection to the Bartlett
Decomposition}\label{connection-to-the-bartlett-decomposition}
\addcontentsline{toc}{subsection}{Connection to the Bartlett
Decomposition}

The default \(d_i = p - i + 1\) has a natural statistical interpretation
through the \textcite{bartlett1934theory} decomposition. If
\(W \sim \mathcal{W}_p(I_p, n)\), the Cholesky factor \(C\) satisfying
\(W = CC^\top\) has squared diagonal entries
\(C_{ii}^2 \sim \chi^2_{n-i+1}\) independently, so
\(\mathrm{E}[C_{ii}^2] = n - i + 1\). At the boundary \(n = p\)---the
minimally proper Wishart---these expected values coincide with the
sequence \(d_i = p - i + 1\), anchoring the GCP diagonal to the natural
scales of the positive-definite cone. Moreover, in the Bartlett
representation the off-diagonal entries satisfy
\(C_{ij} \overset{\mathrm{iid}}{\sim} \mathrm{N}(0,1)\) for \(j < i\),
which motivates placing independent standard Gaussian priors on the free
edge parameters \(\theta_{ij}\); doing so induces a prior on
\(Q(\theta) = L(\theta)L(\theta)^\top\) that shares the scale and
correlation structure of a Wishart prior, but is strictly supported on
the submanifold of \(\mathcal{G}\)-sparse positive definite matrices.

\section*{References}\label{references}
\addcontentsline{toc}{section}{References}

\printbibliography[heading=none]

@book{amari2016information,
  title = {Information {{Geometry}} and {{Its Applications}}},
  author = {Amari, Shun-ichi},
  year = 2016,
  series = {Applied {{Mathematical Sciences}}},
  volume = {194},
  publisher = {Springer Japan},
  address = {Tokyo},
  doi = {10.1007/978-4-431-55978-8},
  urldate = {2026-04-29},
  copyright = {https://www.springernature.com/gp/researchers/text-and-data-mining},
  isbn = {978-4-431-55977-1},
  langid = {english}
}

@article{micha2010red,
  title={Red and processed meat consumption and risk of incident coronary heart disease, stroke, and diabetes mellitus: a systematic review and meta-analysis},
  author={Micha, Renata and Wallace, Sarah K and Mozaffarian, Dariush},
  journal={Circulation},
  volume={121},
  number={21},
  pages={2271--2283},
  year={2010},
  publisher={Lippincott Williams \& Wilkins}
}

@article{li2017sleep,
  title={Sleep duration and smoking are associated with coronary heart disease among US adults with type 2 diabetes: gender differences},
  author={Li, Lixin and Gong, Shaoqing and Xu, Chun and Zhou, Joseph Yi and Wang, Ke-Sheng},
  journal={Diabetes research and clinical practice},
  volume={124},
  pages={93--101},
  year={2017},
  publisher={Elsevier}
}

@article{kong2024bayesian,
  title={Bayesian network analysis of factors influencing type 2 diabetes, coronary heart disease, and their comorbidities},
  author={Kong, Danli and Chen, Rong and Chen, Yongze and Zhao, Le and Huang, Ruixian and Luo, Ling and Lai, Fengxia and Yang, Zihua and Wang, Shuang and Zhang, Jingjing and others},
  journal={BMC Public Health},
  volume={24},
  number={1},
  pages={1267},
  year={2024},
  publisher={Springer}
}

@article{rossouw1983coronary,
title={Coronary risk factor screening in three rural communities. The CORIS
 baseline study},
author={Rossouw, JE and Du Plessis, JP and Benadé, AJ and Jordaan, PC and
 Kotze, JP and Jooste, PL and Ferreira, JJ},
 journal={South African Medical Journal},
volume={64},
number={12},
 pages={430--436},
 year={1983}
 }

@article{atay2005monte,
  title = {A {{Monte Carlo}} Method for Computing the Marginal Likelihood in Nondecomposable {{Gaussian}} Graphical Models},
  author = {{Atay-Kayis}, Aliye and Massam, H{\'e}l{\`e}ne},
  year = 2005,
  journal = {Biometrika},
  volume = {92},
  number = {2},
  pages = {317--335},
  issn = {1464-3510, 0006-3444},
  doi = {10.1093/biomet/92.2.317},
  urldate = {2026-06-10},
  langid = {english}
}

@article{barnard2000modeling,
  title = {Modeling Covariance Matrices in Terms of Standard Deviations and Correlations, with Application to Shrinkage},
  author = {Barnard, John and McCulloch, Robert and Meng, Xiao-Li},
  year = 2000,
  journal = {Statistica Sinica},
  volume = {10},
  number = {4},
  pages = {1281--1311}
}

@article{bartlett1934theory,
  title = {On the {{Theory}} of {{Statistical Regression}}},
  author = {Bartlett, M. S.},
  year = 1934,
  journal = {Proceedings of the Royal Society of Edinburgh},
  volume = {53},
  pages = {260--283},
  issn = {0370-1646},
  doi = {10.1017/S0370164600015637},
  urldate = {2026-05-24},
  copyright = {https://www.cambridge.org/core/terms},
  langid = {english}
}

@article{bekker2017wishart,
  title = {Wishart Distributions: {{Advances}} in Theory with {{Bayesian}} Application},
  shorttitle = {Wishart Distributions},
  author = {Bekker, Andri{\"e}tte and Van Niekerk, Janet and Arashi, Mohammad},
  year = 2017,
  journal = {Journal of Multivariate Analysis},
  volume = {155},
  pages = {272--283},
  publisher = {Elsevier},
  urldate = {2026-05-24}
}

@book{boothby2003introduction,
  title = {An Introduction to Differentiable Manifolds and {{Riemannian}} Geometry},
  author = {Boothby, William M.},
  year = 2003,
  edition = {Rev. 2nd ed},
  publisher = {Academic Press},
  address = {Amsterdam ; New York},
  isbn = {978-0-12-116051-7},
  lccn = {QA614.3 .B66 2003}
}

@article{carpenter2017stan,
  title = {{\emph{Stan}}: {{A}} Probabilistic Programming Language},
  shorttitle = {{\emph{Stan}}},
  author = {Carpenter, Bob and Gelman, Andrew and Hoffman, Matthew D. and Lee, Daniel and Goodrich, Ben and Betancourt, Michael and Brubaker, Marcus and Guo, Jiqiang and Li, Peter and Riddell, Allen},
  year = 2017,
  journal = {Journal of Statistical Software},
  volume = {76},
  number = {1},
  issn = {1548-7660},
  doi = {10.18637/jss.v076.i01},
  urldate = {2024-11-02},
  langid = {english}
}

@article{daniels1999nonconjugate,
  title = {Nonconjugate {{Bayesian Estimation}} of {{Covariance Matrices}} and Its {{Use}} in {{Hierarchical Models}}},
  author = {Daniels, Michael J. and Kass, Robert E.},
  year = 1999,
  journal = {Journal of the American Statistical Association},
  volume = {94},
  number = {448},
  pages = {1254--1263},
  issn = {0162-1459, 1537-274X},
  doi = {10.1080/01621459.1999.10473878},
  urldate = {2026-05-24},
  langid = {english}
}

@book{davis2006direct,
  title = {Direct {{Methods}} for {{Sparse Linear Systems}}},
  author = {Davis, Timothy A.},
  year = 2006,
  publisher = {{Society for Industrial and Applied Mathematics}},
  doi = {10.1137/1.9780898718881},
  urldate = {2026-05-23},
  isbn = {978-0-89871-613-9},
  langid = {english}
}

@article{dempster1972covariance,
  title = {Covariance {{Selection}}},
  author = {Dempster, A. P.},
  year = 1972,
  journal = {Biometrics},
  volume = {28},
  number = {1},
  eprint = {2528966},
  eprinttype = {jstor},
  pages = {157},
  issn = {0006341X},
  doi = {10.2307/2528966},
  urldate = {2026-04-29}
}

@article{drton2017structure,
  title = {Structure {{Learning}} in {{Graphical Modeling}}},
  author = {Drton, Mathias and Maathuis, Marloes H.},
  year = 2017,
  journal = {Annual Review of Statistics and Its Application},
  volume = {4},
  number = {1},
  pages = {365--393},
  issn = {2326-8298, 2326-831X},
  doi = {10.1146/annurev-statistics-060116-053803},
  urldate = {2026-05-24},
  langid = {english}
}

@article{freni2025graphical,
  title = {A Graphical Framework for Interpretable Correlation Matrix Models for Multivariate Regression},
  author = {{Freni-Sterrantino}, Anna and Rustand, Denis and Van Niekerk, Janet and Krainski, Elias and Rue, H{\aa}vard},
  year = 2025,
  journal = {Statistical Methods \& Applications},
  volume = {34},
  number = {3},
  pages = {409--447},
  issn = {1618-2510, 1613-981X},
  doi = {10.1007/s10260-025-00788-y},
  urldate = {2025-08-18},
  langid = {english}
}

@article{gabry2019visualization,
  title = {Visualization in {{Bayesian Workflow}}},
  author = {Gabry, Jonah and Simpson, Daniel and Vehtari, Aki and Betancourt, Michael and Gelman, Andrew},
  year = 2019,
  journal = {Journal of the Royal Statistical Society Series A: Statistics in Society},
  volume = {182},
  number = {2},
  pages = {389--402},
  issn = {0964-1998, 1467-985X},
  doi = {10.1111/rssa.12378},
  urldate = {2026-06-06},
  copyright = {https://academic.oup.com/journals/pages/open\_access/funder\_policies/chorus/standard\_publication\_model},
  langid = {english}
}

@manual{krainski2026graphpcor,
  type = {R Package Version 0.1.25},
  title = {{{graphpcor}}: {{Models}} for Correlation Matrices Based on Graphs},
  author = {Krainski, Elias Teixeira and Rustand, Denis and {Freni-Sterrantino}, Anna and {van Niekerk}, Janet and Rue, H{\aa}vard},
  year = 2026,
  doi = {10.32614/CRAN.package.graphpcor}
}

@book{lauritzen1996graphical,
  title = {Graphical {{Models}}},
  author = {Lauritzen, Steffen L},
  year = 1996,
  publisher = {Oxford University PressOxford},
  doi = {10.1093/oso/9780198522195.001.0001},
  urldate = {2026-05-23},
  isbn = {978-0-19-852219-5},
  langid = {english}
}

@article{lewandowski2009generating,
  title = {Generating Random Correlation Matrices Based on Vines and Extended Onion Method},
  author = {Lewandowski, Daniel and Kurowicka, Dorota and Joe, Harry},
  year = 2009,
  journal = {Journal of Multivariate Analysis},
  volume = {100},
  number = {9},
  pages = {1989--2001},
  issn = {0047-259X},
  doi = {10.1016/j.jmva.2009.04.008},
  urldate = {2026-03-24}
}

@book{murray1993differential,
  title = {Differential Geometry and Statistics},
  author = {Murray, M. K. and Rice, John W.},
  year = 1993,
  series = {Monographs on Statistics and Applied Probability},
  edition = {1st ed},
  number = {48},
  publisher = {Chapman \& Hall},
  address = {London ; New York},
  isbn = {978-0-412-39860-5},
  lccn = {QA276 .M85 1993}
}

@article{pymc2023,
  title = {{{PyMC}}: {{A}} Modern and Comprehensive Probabilistic Programming Framework in {{Python}}},
  author = {{Abril-Pla}, Oriol and Andreani, Virgile and Carroll, Colin and Dong, Larry and Fonnesbeck, Christopher J. and Kochurov, Maxim and Kumar, Ravin and Lao, Junpeng and Luhmann, Christian C. and Martin, Osvaldo A. and Osthege, Michael and Vieira, Ricardo and Wiecki, Thomas and Zinkov, Robert},
  year = 2023,
  journal = {PeerJ Computer Science},
  volume = {9},
  number = {e1516},
  doi = {10.7717/peerj-cs.1516}
}

@article{rose1976algorithmic,
  title = {Algorithmic {{Aspects}} of {{Vertex Elimination}} on {{Graphs}}},
  author = {Rose, Donald J. and Tarjan, R. Endre and Lueker, George S.},
  year = 1976,
  journal = {SIAM Journal on Computing},
  volume = {5},
  number = {2},
  pages = {266--283},
  issn = {0097-5397, 1095-7111},
  doi = {10.1137/0205021},
  urldate = {2026-05-23},
  langid = {english}
}

@article{roverato2002hyper,
  title = {Hyper {{Inverse Wishart Distribution}} for {{Non}}-decomposable {{Graphs}} and Its {{Application}} to {{Bayesian Inference}} for {{Gaussian Graphical Models}}},
  author = {Roverato, Alberto},
  year = 2002,
  journal = {Scandinavian Journal of Statistics},
  volume = {29},
  number = {3},
  pages = {391--411},
  issn = {0303-6898, 1467-9469},
  doi = {10.1111/1467-9469.00297},
  urldate = {2026-06-10},
  langid = {english}
}

@book{rue2005gaussian,
  title = {Gaussian {{Markov}} Random Fields: {{Theory}} and Applications},
  shorttitle = {Gaussian {{Markov Random Fields}}},
  author = {Rue, H{\aa}vard and Held, Leonhard},
  year = 2005,
  publisher = {{Chapman and Hall/CRC}},
  address = {New York},
  doi = {10.1201/9780203492024},
  isbn = {978-0-429-20882-9}
}

@article{rue2009approximate,
  title = {Approximate {{Bayesian}} Inference for Latent {{Gaussian}} Models by Using Integrated Nested {{Laplace}} Approximations},
  author = {Rue, H{\aa}vard and Martino, Sara and Chopin, Nicolas},
  year = 2009,
  journal = {Journal of the Royal Statistical Society Series B: Statistical Methodology},
  volume = {71},
  number = {2},
  pages = {319--392},
  issn = {1369-7412, 1467-9868},
  doi = {10.1111/j.1467-9868.2008.00700.x},
  urldate = {2023-11-27},
  langid = {english}
}

@article{simpson2017penalising,
  title = {Penalising Model Component Complexity: {{A}} Principled, Practical Approach to Constructing Priors},
  shorttitle = {Penalising Model Component Complexity},
  author = {Simpson, Daniel and Rue, H{\aa}vard and Riebler, Andrea and Martins, Thiago G. and S{\o}rbye, Sigrunn H.},
  year = 2017,
  journal = {Statistical Science},
  volume = {32},
  number = {1},
  pages = {1--28},
  publisher = {Institute of Mathematical Statistics},
  issn = {0883-4237, 2168-8745},
  doi = {10.1214/16-STS576},
  urldate = {2025-09-05}
}

@article{speed1986gaussian,
  title = {Gaussian {{Markov Distributions}} over {{Finite Graphs}}},
  author = {Speed, T. P. and Kiiveri, H. T.},
  year = 1986,
  journal = {The Annals of Statistics},
  volume = {14},
  number = {1},
  issn = {0090-5364},
  doi = {10.1214/aos/1176349846},
  urldate = {2026-06-05}
}

@article{tokuda2025visualizing,
  title = {Visualizing Distributions of Covariance Matrices},
  author = {Tokuda, Tomoki and Goodrich, Ben and Van Mechelen, Iven and Gelman, Andrew and Tuerlinckx, Francis},
  year = 2025,
  journal = {Journal of Data Science, Statistics, and Visualisation},
  volume = {5},
  number = {7},
  urldate = {2026-05-24}
}

@article{wermuth1980linear,
  title = {Linear {{Recursive Equations}}, {{Covariance Selection}}, and {{Path Analysis}}},
  author = {Wermuth, Nanny},
  year = 1980,
  journal = {Journal of the American Statistical Association},
  volume = {75},
  number = {372},
  pages = {963--972},
  issn = {0162-1459, 1537-274X},
  doi = {10.1080/01621459.1980.10477580},
  urldate = {2026-06-10},
  langid = {english}
}

@article{wishart1928generalised,
  title = {The Generalised Product Moment Distribution in Samples from a Normal Multivariate Population},
  author = {Wishart, John},
  year = 1928,
  journal = {Biometrika},
  volume = {20},
  number = {1/2},
  eprint = {2331939},
  eprinttype = {jstor},
  pages = {32--52},
  publisher = {JSTOR},
  urldate = {2026-05-24}
}

@article{yannakakis1981computing,
  title = {Computing the {{Minimum Fill-In}} Is {{NP-Complete}}},
  author = {Yannakakis, Mihalis},
  year = 1981,
  journal = {SIAM Journal on Algebraic Discrete Methods},
  volume = {2},
  number = {1},
  pages = {77--79},
  issn = {0196-5212, 2168-345X},
  doi = {10.1137/0602010},
  urldate = {2026-05-23},
  langid = {english}
}

\end{document}